\documentclass[usenatbib]{mnras}
\usepackage{txfonts}
\usepackage[utf8x]{inputenc}
\usepackage{graphicx}
\usepackage{natbib}
\usepackage{url}
\usepackage{amssymb}
\newcommand{\HI}{H\,{\sevensize I}}
\newcommand{\cii}{ [C\,{\sevensize II}]~158~$\mu$m}
\newcommand{\CpHt}{${\rm C^+/H_2}$}
\title{C$^+$/H$_2$ Gas in Star-Forming Clouds and Galaxies}

\author[Nordon \& Sternberg]{
Raanan Nordon,$^{1}$\thanks{nordon@astro.tau.ac.il}
Amiel~Sternberg,$^{1}$
\\
$^{1}$School of Physics and Astronomy, Faculty of Exact Sciences, Tel-Aviv University, Tel-Aviv 69978, Israel. \\
}

\begin{document}
\maketitle

\begin{abstract}
We present analytic theory for the total column density of singly ionized carbon (C$^+$) in the optically thick photon dominated regions (PDRs) of far-UV irradiated (star-forming) molecular clouds.
We derive a simple formula for the C$^+$ column as a function of the cloud (hydrogen) density, the far-UV field intensity, and metallicity, encompassing the wide range of galaxy conditions.
When assuming the typical relation between UV and density in the cold neutral medium, the C$^+$ column becomes a function of the metallicity alone.
We verify our analysis with detailed numerical PDR models.
For optically thick gas, most of the C$^+$ column is mixed with hydrogen that is primarily molecular (H$_2$), and this ``\CpHt'' gas layer accounts for almost all of the `CO-dark' molecular gas in PDRs.
The \CpHt\ column density is limited by dust shielding and is inversely proportional to the metallicity down to $\sim$0.1 solar.
At lower metallicities, H$_2$ line blocking dominates and the \CpHt\ column saturates.
Applying our theory to CO surveys in low redshift spirals we estimate the fraction of \CpHt\ gas out of the total molecular gas to be typically $\sim$0.4.
At redshifts $1<z<3$ in massive disc galaxies the \CpHt\ gas represents a very small fraction of the total molecular gas ($\lesssim0.16$).
This small fraction at high redshifts is due to the high gas surface densities when compared to local galaxies.
\end{abstract}

\begin{keywords}
galaxies:ISM:clouds -- ISM:molecules -- ISM:photodissociation region (PDR) -- galaxies:ISM -- submillimetre: ISM
\end{keywords}

%%%%%%%%%%%%%%%%%%%%%%%%%%%%%%%%%%%%%%%%%%%%%%%%
%
\section{Introduction} \label{sec:Introduction}
%
%%%%%%%%%%%%%%%%%%%%%%%%%%%%%%%%%%%%%%%%%%%%%%%%

Gas in galaxies can be directly observed by several methods.
Most commonly and accessibly, the atomic \HI\ gas is observed in its 21~cm emission and molecular H$_2$ is observed through tracer molecules - mainly CO rotational lines.
But do these tracers account for all the neutral and molecular gas?
From models of the structure of molecular clouds one should expect a region in the cloud in which hydrogen is predominantly in H$_2$ form but the carbon exists mostly as C$^+$ and has not formed CO.
Such C$^+$/H$_2$ gas emits neither in 21~cm nor in CO lines.

Where will such a region exist in a cloud?
In order to maintain the gas as molecular in H$_2$ and CO, both must be shielded from dissociation by UV photons at wavelengths close to their spectral absorption lines.
This shielding is achieved by a combination of dust, and H$_2$ \& CO self and mutual shielding.
However, the much more abundant H$_2$ is more efficient at shielding itself than at shielding CO
%due to miss-alignment of the spectral lines
and thus, the \HI$\to$H$_2$ transition occurs before CO is fully shielded by dust, H$_2$ and itself \citep{vanDishoeck1988, Sternberg2014, Bialy2016}.
In these regions UV photons can still pass between the H$_2$ lines and ionize the carbon.
The gas that exists deeper than the \HI$\to$H$_2$ transition but before CO is fully shielded is the C$^+$/H$_2$ gas and is sometimes referred to as `dark gas'.
Of course this term is a misnomer since this gas does emit in other spectral lines, most relevant of which is \cii.
We will refer to this gas as \CpHt, or `CO-dark gas'.
A population of neutral C also exists outside of the CO dominated region though its column is smaller than the C$^+$ in most relevant conditions. 

Indeed, observational evidence indicate to the presence of non-negligible amounts of C$^+$/H$_2$ gas in local clouds and overall in galaxies.
\citet{Reach1994} concluded from resolved \HI\ and CO maps of molecular clouds that the H$_2$ must be spatial more extent than the CO-detected regions.
\citet{Grenier2005} used gamma rays emitted by interaction of cosmic-rays (CR) with the gas to measure the total column density in the solar neighborhood and compared that to the columns measured from \HI\ and CO maps. They inferred a significant column of `CO-dark gas' of the same order as the molecular gas detected through CO lines.
\citet{Langer2010} found excess of C$^+$ emissions from diffuse clouds and concluded that the excess must be attributed to a molecular phase with no CO in it.
More recently, \citet{Pineda2013} and \citet{Langer2014} used far-infrared and sub-millimeter spectroscopy from {\it Herschel} to study the molecular content of the Milky Way.
They concluded that 30--40\% of the Milky Way molecular gas is in the \CpHt\ phase, and that the fraction in individual clouds depends on the column density of the cloud - higher fractions in diffuse clouds and lower in dense clouds.

In extragalactic observations the total gas column can be indirectly inferred from modeling the the far-infrared emission to derive the dust content and converting the dust column into gas column.
\citet{Israel1997} and \citet{Leroy2007} used such dust modeling methods in low metallicity galaxies and conclude that large amounts of H$_2$ gas is not detected in CO.
\citet{Leroy2011} extended the study to a range of metallicities and find that the inferred amount of the `CO-dark gas' increase with decreasing metallicity.
\citet{Genzel2012} reached a similar conclusion based on CO measurements and arguments of star formation efficiency.

The amount and nature of the gas which is not detected in either \HI\ or CO is somewhat disputed as some authors claim that part, if not most of this gas is atomic rather than molecular, and is undetected in \HI\ as the line becomes optically thick \citep{Braun2009, Fukui2015}.
In this paper we focus on the \CpHt\ gas that is `CO dark' and by definition is in the molecular phase, and also consider the C$^+$ in the \HI\ region of clouds that are thick enough to include a CO core.

\citet[][W10 hereafter]{Wolfire2010} modeled molecular clouds as a sphere with a $r^{-1}$ density profile and used a numerical code to calculate the abundance profiles inside such a cloud.
Using this model they studied the fraction of the `CO-dark gas' as a function of various parameters such as the radiation field, metallicity and mean surface density of the cloud.
They estimate a typical `CO-dark gas' fraction of $\sim$0.3 in giant molecular clouds (GMCs), where the strongest dependency is on the mean total optical depth (from one side to the other) of the cloud.
Their result is in a generals agreement with the 3D hydro simulations of \citet{Glover2010}.

The main goal of this paper is to derive a simple analytic formula to the total column of the C$^+$/H$_2$ gas, based on first principles, and then test the analytic results against a numerical code.
In Section~\ref{sec:models}, we describe our simple model of a plane-parallel cloud.
In Section~\ref{sec:analytic_C+_column}, we present our analytic formula of the C$^+$ column and compare it with the numerical models.
%In Section~\ref{sec:carbon_chem} we discuss the carbon chemistry in molecular clouds.
In Section~\ref{sec:numerical_models}, we describe the grid of numerical models we calculated in order to test the effects of different parameters.
In Section~\ref{sec:neutral_C}, we discuss the column of neutral C between the C$^+$ and CO dominated regions.
In Sections~\ref{sec:C+/H2_gas} and \ref{sec:CII_emission} we discuss the expected \CpHt columns in galaxies and the \cii\ emission from this gas.
Finally, in Section~\ref{sec:Discussion} we discuss our results in the context of CO surveys and the total gas content of galaxies.

\begin{table*}[t]
 \caption{Summary of the symbols used in this paper.}
 \label{tab:symbols}
\centering
 \begin{tabular}{c|c|c|l}
        &         & Units/   & \\
 Symbol & Meaning & Value & Comments\\
\hline\hline
$I_{\rm UV}$ & UV radiation density & ... & Relative to isotropic Draine field \\
$\Gamma$ & C photoionization rate& 3$\times10^{-10}$ s$^{-1}$ & in UV field $I_{\rm UV}=1$ \\
$Z^\prime$ & metallicity & ... & relative to Solar\\
$A$ & solar carbon abundance & 1.32$\times 10^{-4}$ & Gas-phase C/H number density ratio at $Z^\prime=1$\\
$n$ & hydrogen density & cm$^{-3}$ & Total hydrogen nucleons \\
$N$ & total hydrogen column & cm$^{-2}$ & measured from one face of the cloud\\
$N_{\rm X}$ & column of specie X & cm$^{-2}$ & $X={\rm HI,H_2,C^+,C^0,CO}$\\
$N_{\rm (HI+H_2), C^+}^{\rm tot}$ & hydrogen gas column associated with C$^+$ & cm$^{-2}$ & $N_{\rm C^+}^{\rm tot}/(AZ^\prime)$ \\
$N_{\rm C^+/H_2}$ & total hydrogen column of the ${\rm C^+/H_2}$ gas & cm$^{-2}$ & Same as $N_{\rm (HI+H_2), C^+}^{\rm tot}$ when neglecting the \HI\ gas \\
$\sigma_{\rm g,\odot}$ & dust absorption cross-section & $1.77\times10^{-21}$~cm$^{2}$ & per hydrogen nucleon at $Z^\prime=1$\\
$\sigma_{\rm H_2}$ & effective H$_2$ line-blocking cross-section & cm$^{2}$ & a function of the H$_2$ column\\

\hline
\end{tabular} 
\end{table*}

%%%%%%%%%%%%%%%%%%%%%%%%%%%%%%%%%%%%%%%%%%%%%%%%%%%%%%%%%%%%%
%
\section{Models} \label{sec:models}
%
%%%%%%%%%%%%%%%%%%%%%%%%%%%%%%%%%%%%%%%%%%%%%%%%%%%%%%%%%%%%%

\begin{figure}[t]
 \centering
 \includegraphics[width=\columnwidth, clip=true, trim=0pt 0pt 0cm 0cm]{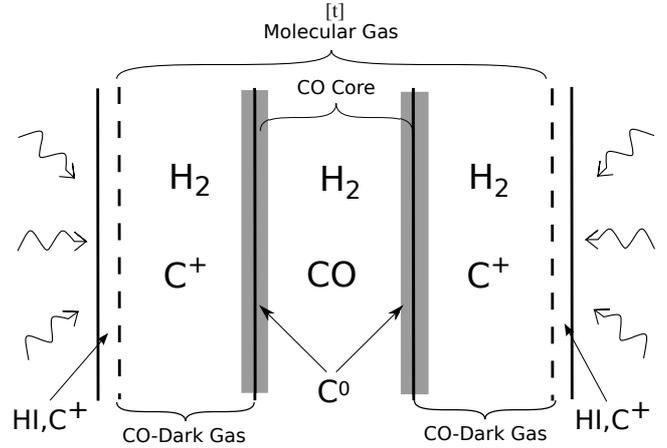}
 \caption{Schematic structure of an optically thick slab illuminated from both sides, showing the ${\rm C^+/\textrm{H\,\sevensize{I}}}$, ${\rm C^+/H_2}$, ${\rm C/H_2}$, ${\rm CO/H_2}$ zones}
 \label{fg:cloud_sketch}
\end{figure}

We consider the standard \HI/H$_2$ and C$^+$/C/CO transitions and chemical structures for plane-parallel photon dominated regions (PDRs), as illustrated in Figure~\ref{fg:cloud_sketch}.
We focus on optically thick slabs illuminated on both sides by an isotropic far-UV photodissociating and photoionizing radiation.
In general, the \HI\ to H$_2$ transition occurs at smaller cloud depths compared to the location of the C$^+$/C/CO transitions \citep{Tielens1985a, Sternberg1995}.
The cloud depth from one cloud surface to the mid-plane may be parametrized by the total hydrogen gas column density $N \equiv N_{\rm HI} + 2N_{\rm H2}$, or by the associated and metallicity-dependent visual extinction $A_{\rm V}$ into the cloud.
Here, $N_{\rm HI}$ and $N_{\rm H_2}$ are the columns of atomic and molecular hydrogen from the surface to the given depth.
For optically thick slabs (Figure~\ref{fg:cloud_sketch}) the radiative transfer and chemical structures on each side are equivalent to that in a semi-infinite slab illuminated from one side.

The structure consists of four zones.
First is an outer ``${\textrm C^+/\textrm{H\,\sevensize{I}}}$'' layer in which the carbon is ionized and the hydrogen is atomic.
Second is the ``\CpHt'' layer in which the carbon is still ionized, but the hydrogen is molecular.
Third is ``${\rm C/H_2}$'' in which the carbon is in neutral atomic form.
Fourth is the ``${\rm CO/H_2}$'' core in which most of the carbon is locked in CO (and the hydrogen is H$_2$).
The relative sizes of these zones depend in general on several parameters, including the gas density, the far-UV intensity, the CR ionization rate, and the gas-phase oxygen and carbon abundances, as set by the metallicity and grain depletion factors.

In \S~\ref{sec:numerical_models} we present numerical computations for the C$^+$/C/CO structures for a wide range of metallicities.
As we will show, also analytically in \S~\ref{sec:analytic_C+_column}, the ${\rm C^+/\textrm{H\,\sevensize{I}}}$ layer is generally small or negligible compared to \CpHt, and the ${\rm C/H_2}$ layer is also small.
Thus, the \CpHt\ column represents the ``CO-dark'' H$_2$ gas that is not mixed with CO.
Our analytic formula (\S~\ref{sec:analytic_C+_column}) for the C$^+$ column density and associated \CpHt\ hydrogen gas column provides a convenient and simple estimate for the CO-dark H$_2$ mass.

%-------------------------------
\subsection[The H I Column Density]{The H\,{\sevensize I} column density}
\label{sec:HI_column}
%-------------------------------
\citet{Sternberg2014} derived the analytic formula
\begin{equation}
 N_{\rm HI}^{\rm tot} = \frac{\langle\mu\rangle}{\sigma_{\rm g}} \ln\left(\frac{1}{4} \frac{\alpha G}{\langle\mu\rangle} +1\right) \\\ ,
\label{eq:N_HI_total}
\end{equation}
for the total \HI\ column density produced by far-UV photodissociation on one side of optically thick clouds illuminated by isotropic radiation fields.
In this expression, $\langle\mu\rangle=0.8$ is an average angle factor that appears for irradiation by isotropic fields \citep{Sternberg2014}.
The dimensionless parameter is
\begin{equation}
 \alpha G \equiv \frac{D_0}{Rn}G = \bar{f}_{\rm diss} \frac{\sigma_{\rm g} w F_{\rm LW}}{Rn} \\\ .
 \label{eq:alpha_G_definition}
\end{equation}
Here, $n=n_1 + 2n_2$ is the total (atomic plus molecular) hydrogen gas density (cm$^{-3}$), $R$ is the dust grain H$_2$ formation rate coefficient (cm$^3$~s$^{-1}$), $D_0$ is the optically thin H$_2$ dissociation rate (s$^{-1}$) in a ``free-space'' radiation field, $G$ is the average H$_2$ self-shielding factor.
The numerator can also be expressed using  $w F_{\rm LW}$ that is the effective photon flux (cm$^{-2}$~s$^{-1}$) of Lyman-Werner (LW) band radiation available for H$_2$ dissociation in dusty clouds,
$\bar{f}_{\rm diss} = 0.12$ is the mean dissociation probability averaged over all lines,
and $\sigma_{\rm g}$ is the mean dust absorption cross-section per hydrogen nucleus within the LW band.
For a radiation field normalized to the \citet{Draine1978} representation of the Galactic interstellar field $D_0=5.8\times10^{-11} I_{\rm UV}$~s$^{-1}$ and $F_{\rm LW}=2.07\times10^{7}I_{\rm UV}$~cm$^{-2}$~s$^{-1}$, where $I_{\rm UV}$ is the field intensity factor.
The parameter $w\leq1$ is the normalized H$_2$-dust limited LW absorption bandwidth.
For solar metallicity and typical dust-to-gas mass ratios, $w\sim0.5$.

If the dust-to-gas ratio is proportional to the metallicity then the mean dust absorption cross section per H nucleon for LW-band photons is
\begin{equation}
  \sigma_g = \sigma_{\rm g,\odot} Z^\prime = 1.9\times10^{-21} \phi_g Z^\prime \\\ .
  \label{eq:sigma_g_def}
\end{equation}
We parametrize the metallicity by $Z^\prime$, where $Z^\prime=1$ corresponds to solar photospheric abundances of the heavy elements.
For solar metallicity $\sigma_{\rm g} = \sigma_{\rm g,\odot}$.
In Eq.~\ref{eq:sigma_g_def}, $\phi_g$ is a factor of order unity that depends on the dust absorption and scattering properties.
Here we adopt $\phi_g=0.93$.

In terms of the UV-field intensity, gas density, and metallicity
\begin{equation}
 \alpha G = 1.54 \frac{\phi_g}{1+(2.64 \phi_g Z^\prime)^{1/2}} \frac{I_{\rm UV}}{(n/100\,{\rm cm^{-3}})} \\\ .
 \label{eq:alpha_G}
 \end{equation}
For $\alpha G>1$ (strong field limit) absorption of LW-band photons is dominated by dust associated with an extended outer \HI\ layer.
When $\alpha G<1$ (weak field limit) the absorption of LW-band photons is dominated by H$_2$ lines and just the dust associated with the H$_2$.
In general, $\alpha G \propto I_{\rm UV}/n$, and can range from small to large.

For a self-regulated medium \citep{Wolfire2003, McKee2010, Ostriker2010} in which the \HI\ gas is driven to densities and thermal pressures allowing multiphase warm and cold neutral medium (WNM/CNM) gas, and heated by photoelectric emission from dust grains, the gas density is $n\approx n_{\rm CNM}$ where
\begin{equation}
 n_{\rm CNM} \approx \frac{93}{1+3.1Z^{\prime 0.365}} I_{\rm UV} \quad {\rm cm}^{-3} \\\ .
 \label{eq:CNM_relation}
\end{equation}
This relationship between $n$ and $I_{\rm UV}$ then restricts $\alpha G$ to a narrow range of order unity,
\begin{equation}
 \alpha G \approx (\alpha G)_{\rm CNM} = 1.66 \phi_g \frac{1+3.1Z^{\prime 0.365}}{1+1.62(\phi_g Z^\prime)^{1/2}} \\\ ,
\end{equation}
weakly dependent on the metallicity.

%%%%%%%%%%%%%%%%%%%%%%%%%%%%%%%%%%%%%%
%
\section{Analytic Formula for C$^+$ Column Density} \label{sec:analytic_C+_column}
%
%%%%%%%%%%%%%%%%%%%%%%%%%%%%%%%%%%%%%%

%---------------------------------------
\subsection{C$^+$ Chemistry} \label{sec:carbon_chem}
%---------------------------------------

In PDRs, C$^+$ is produced primarily by photoionization
\begin{equation}
 {\rm C} + \nu \to {\rm C}^+ + {\rm e}^- \\\ ,
\end{equation}
by photons with energies between the neutral carbon ionization threshold 11.26~eV and the Lyman limit at 13.6~eV.
Remarkably, this energy range corresponds almost exactly to the LW 11.2--13.6~eV band for H$_2$-line dissociation.
The continuum carbon photoionization cross section is $\sigma_{\rm C} = 1.5\times10^{-17}$ cm$^2$ \citep{Dischoeck2006}, and the photoionization rate in a radiation field with $I_{\rm UV}=1$ is then $\Gamma = F_{\rm LW} \sigma_{\rm C} = 3\times10^{-10}$ s$^{-1}$.

There are three dominant processes for the removal of the C$^+$.
First is neutralization via radiative recombination
\begin{equation}
 {\rm C}^+ + {\rm e}^- \to {\rm C} + \nu \\\ .
\end{equation}
The recombination rate coefficient, radiative plus dielectronic \citep{Altun2004, Badnell2006, Wolfire2008} may be expressed as
\begin{equation}
 \alpha_{\rm e} = 1.8\times10^{-11} \left( \frac{T}{100\,{\rm K}} \right)^{-0.83} \quad {\rm cm^{3}\:s^{-1}} \\\ ,
\end{equation}
where $T$ is the gas temperature.

Second is grain-assisted neutralization,
\begin{equation}
 {\rm C^+ + e:gr \to C + gr} \quad .
\end{equation}
The rate coefficient for this process depends on the grain size distribution, charge, and the gas temperature.
For an MRN grain size distribution \citep[][]{Mathis_Rumpl_Nordsieck1977} we can use the \citet[][]{Draine_Sutin1987} expression for the rate coefficient per hydrogen nucleon (their Eq.~5.15).
For C$^+$ and for a solar metallicity dust-to-gas ratio
\begin{equation}
 \alpha_{\rm gr} \approx 2.9\times10^{-15} \left(\frac{a_{\rm min}}{30~{\rm \AA}}\right)^{-3/2} \left(\frac{T}{100~{\rm K}}\right)^{-1/2} \quad {\rm cm^{3}\,s^{-1}} \\\ ,
\end{equation}
where $a_{\rm min}$ is the minimum grain size.
We adopt a standard $a_{\rm min} = 30$~\AA\ and $T=100$~K.
The dust neutralization rate coefficient scales linearly with the dust-to-gas ratio, and we assume that the latter scales linearly with the metallicity $Z^\prime$.

The third removal process is radiative association with H$_2$,
\begin{equation}
 {\rm C}^+ + {\rm H}_2 \to {\rm CH_2^+} + \nu \\\ .
\end{equation}
The reaction rate coefficient \citep{Gerlich1994, KIDA} is
\begin{equation}
 k_{\rm H_2} = 3.3\times10^{-16} \phi_{o/p} \left( \frac{T}{\rm 100\,K} \right)^{-1.30} \exp\left(\frac{\rm -23.0\,K}{T}\right) \quad {\rm cm^3\,s^{-1}} \\\ ,
\end{equation}
where $\phi_{o/p}=1$ for ortho-H$_2$ and $\phi_{o/p}=2.5$ for para-H$_2$.
In our analysis and numerical models (\S~\ref{sec:numerical_models}) we adopt $\phi_{o/p}=1$.

%----------------------------------------------------
\subsection{{\rm C}$^+$ formation-destruction equation} \label{sec:formatio_destruction}
%---------------------------------------------------

With the above processes in mind, a formation-destruction equation for the steady state C$^+$/C density ratio at any cloud depth may be written down.

For a cloud embedded in an isotropic field with $I_{\rm UV}=1$, the photoionization rate at any cloud depth, from the cloud surface is
\begin{equation}
 \Gamma(N) = \frac{\Gamma}{2} \displaystyle\int\displaylimits_{0}^{1} e^{-\tau/\mu} d\mu \\\ .
\end{equation}
Here, $N \equiv N_{\rm \HI}+2N_{\rm H_2}$ is the hydrogen gas column from the surface to the given cloud depth.
$\Gamma$ is the free space ionization rate, $\Gamma/2$ is the rate at the surface of an optically thick cloud, and $\mu \equiv \cos \theta$, where $\theta$ is the angle relative to the normal.
For any ray direction the photoionization rate is reduced by an exponential factor $e^{-\tau/\mu}$, where $\tau$ is the effective opacity (in the normal direction).
The effective opacity is due to the combination of dust absorption and H$_2$-line blocking of the photoionizing radiation.
As we discuss further below, CI opacity is always negligible.

The C$^+$/C formation-destruction equation is then
\begin{equation}
 \left( \alpha_{\rm e}n_{\rm e} + \alpha_{\rm gr}Z^\prime n + k_{\rm H_2} n_{\rm H_2} \right) n_{C^+} \equiv R_{\rm eff}n n_{\rm C^+} = I_{\rm UV} \Gamma(N) n_{\rm C} \\\ ,
 \label{eq:rate_equation}
\end{equation}
where $n_{C^+}$ and $n_{\rm C}$ are the local C$^+$ and C densities (cm$^{-3}$) respectively.
In Eq.~\ref{eq:rate_equation}, we have defined $R_{\rm eff}$ as the total effective C$^+$ neutralization rate coefficient per hydrogen nucleus.

\begin{figure}[t]
 \centering
 \includegraphics[width=\columnwidth, clip=true, trim=0pt 0pt 0pt 20pt]{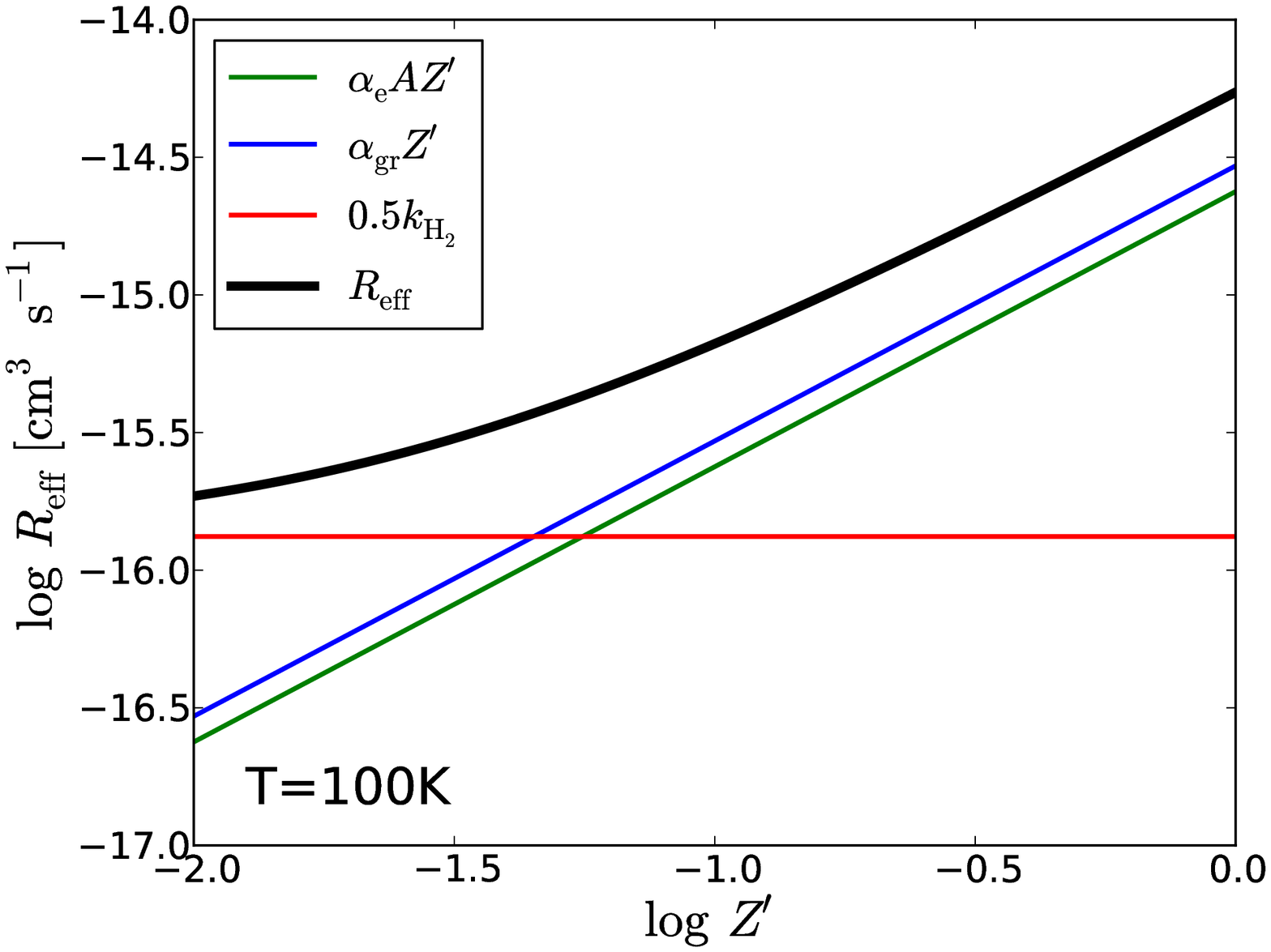} \\
 \includegraphics[width=\columnwidth, clip=true, trim=0pt 0pt 0pt 30pt]{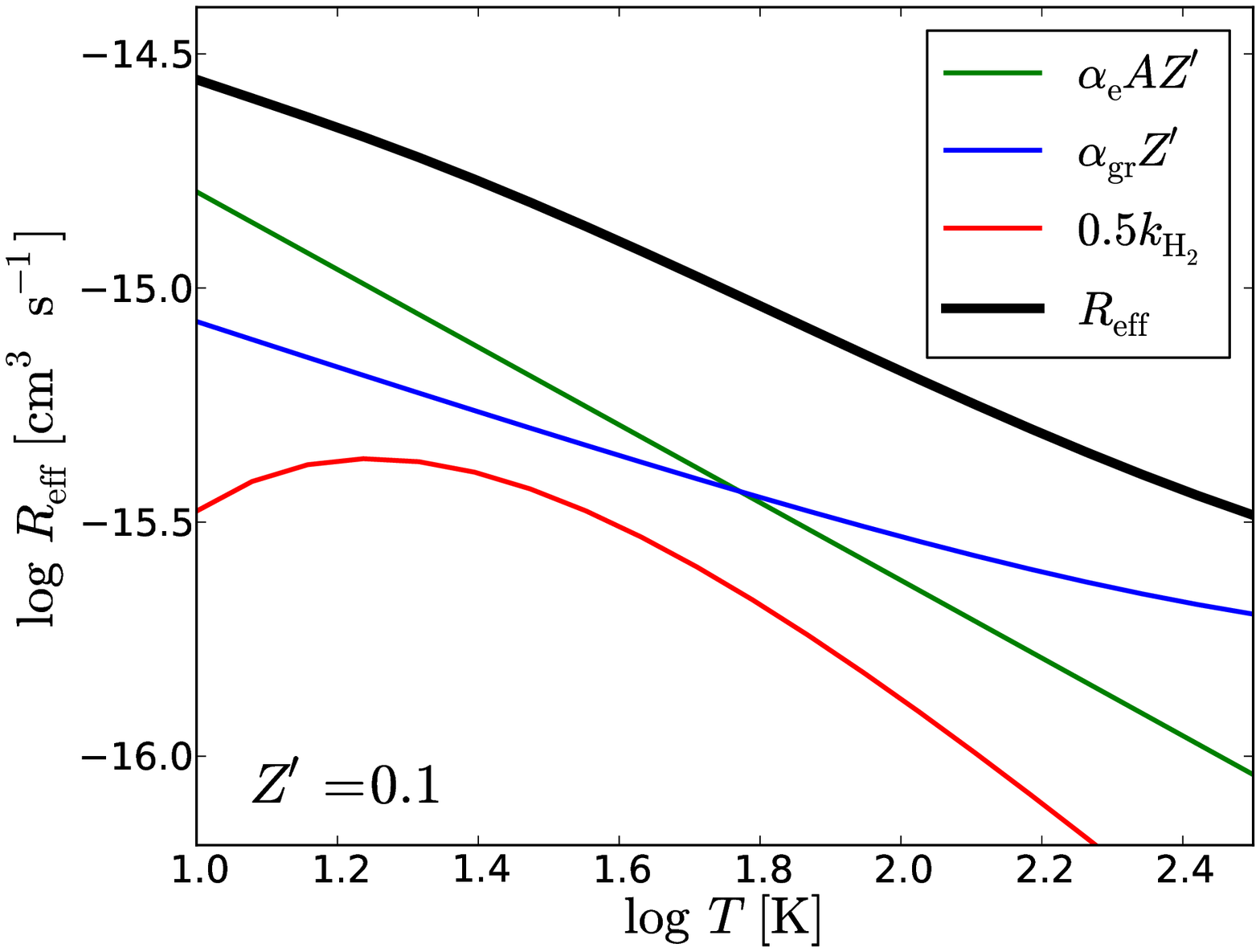}
 % Rate_coeff_Z.eps: 0x0 pixel, 300dpi, 0.00x0.00 cm, bb=13 175 598 616
 \caption{\label{fg:Rate_coeffs} {Top}: the carbon neutralization rate coefficient and components as a function of metalicity at temperature of 100~K.
 {Bottom}: same as above, except as a function of temperature at constant 0.1 solar metallicity.
 }
\end{figure}

To a good approximation, the electron density $n_{\rm e} \approx n_{\rm C^+}$ in the C$^+$ layer.
This approximation holds unless the metallicity is extremely low and/or the CR ionization rate is very high \citep[e.g.,][]{Bialy2015} in which case H$^+$ can become the dominant positive charge carrier as opposed to C$^+$.
Furthermore, since all available carbon is ionized, $n_{\rm C^+}=AZ^\prime n$ where $A$ is the gas phase carbon abundance at solar metallicity.
We set $A=1.32\times10^{-4}$.
We also assume that the hydrogen is fully H$_2$ in the C$^+$ layer so that $n_{\rm H_2}=n/2$.
With these assumptions, the C$^+$ neutralization rate coefficient is
\begin{equation}
 R_{\rm eff} = \alpha_{\rm e} A Z^\prime + \alpha_{\rm gr} Z^\prime + 0.5 k_{\rm H_2} \\\ .
 \label{eq:R_eff}
\end{equation}
In Figure~\ref{fg:Rate_coeffs} we plot $R_{\rm eff}$, as well as its three individual components, as a function of $Z^\prime$ and $T$.

The radiative association rate coefficient $k_{\rm H2}$ does not depend on the metallicity and although for $Z^\prime=1$ it is small compared to the recombination processes, it does become important when the metallicity is low ($Z\lesssim0.1$).
Thus, at high metallicities, $R_{\rm eff} \propto Z^{\prime}$, and at sufficiently low metallicities $R_{\rm eff}$ becomes weakly dependent on $Z^\prime$.
For $a_{\rm min}=30$~\AA, dust neutralization is comparable to radiative recombination.
For a much smaller $a_{\rm min}=3$~\AA\ \citep[][]{Draine_Sutin1987, Lepp1988} dust neutralization can dominate at all temperatures and metallicities.

Typical temperatures in the outer regions of the clouds where carbon is in C$^+$ form are similar to those measured in the HI CNM.
Temperatures are of the order of 100~K, though various sources quote temperatures in the range of 50--200~K either measured \citep[e.g.,][]{Mebold1982, Payne1983, Kulkarni1988, Heiles2003}, or numerically calculated \citep[e.g.,][]{Wolfire2003, Wolfire2010}.
For convenience, we normalize our results to a temperature of 100~K.
However, as we shall see below the resulting C$^+$ column is very weakly dependent on the assumed temperature.

%----------------------------------------
\subsection{Pure Dust Attenuation} \label{sec:pure_dust_attenuation}
%----------------------------------------

We start with the simple high-$Z^\prime$ limit ($Z^\prime \gtrsim 0.1$) for which the carbon-ionizing photons are absorbed by dust, with negligible H$_2$ line-blocking.
We will assume that in the volume where C is photoionized all the carbon is in either C or C$^+$ form, with negligible CO.

For any metallicity the dust optical depth is
\begin{equation}
 \tau \equiv \sigma_{\rm g} N = \sigma_{\rm g,\odot} Z^\prime N = \left(\frac{\sigma_{\rm g,\odot}}{A}\right) ({N}_{\rm C} + {N}_{\rm C+}) \\\ ,
 \label{eq:tau_definition_dust_only}
\end{equation}
where we have assumed that the carbon abundance $AZ^\prime$ and the dust absorption cross section $\sigma_{\rm g,\odot}Z^\prime$ are both linear in $Z^\prime$.
$N_{\rm C^+}$ and $N_{\rm C}$ are the C$^+$ and C columns from the edge of the cloud to a given depth.
The density ratio $n_{\rm C^+}/n_{\rm C} \equiv d{N}_{\rm C^+}/d{N}_{\rm C}$, so the formation destruction equation (Eq.~\ref{eq:rate_equation}) may be written as the differential equation
\begin{equation}
 \frac{dN_{\rm C^+}}{dN_{\rm C}} =
 \frac{1}{2} \mathcal{Y} \displaystyle\int\displaylimits_{0}^{1} e^{ - \left( \sigma_{\rm g,\odot}/A \right) \left( N_{\rm C^+}+N_{\rm C} \right)/\mu } d\mu \,,
\end{equation}
where the dimensionless parameter
\begin{equation}
 \mathcal{Y} \equiv \frac{\Gamma}{R_{\rm eff}} \frac{I_{\rm UV}}{n} \,.
\end{equation}

The parameter $\mathcal{Y}$ for C$^+$/C is analogous to $\alpha G$ for \HI/H$_2$ (\S~\ref{sec:HI_column}).
It is the ratio of the free space carbon photoionization rate to the effective neutralization rate. 
The C$^+$/C density ratio at the cloud surface equals $\mathcal{Y}/2$.
Normalizing to the characteristic value for $R_{\rm eff}$ at $Z^\prime=1$ (see Figure~\ref{fg:Rate_coeffs})
\begin{equation}
 \mathcal{Y} \equiv 545 \left(\frac{\Gamma}{\rm 3\times10^{-10}\,s^{-1}}\right) \left(\frac{\rm 5.5\times10^{-15}\,cm^{3}\,s^{-1}}{R_{\rm eff}}\right) \left(\frac{I_{\rm UV}}{n/100\,{\rm cm^{-3}}}\right) \\\ .
 \label{eq:Y_definition}
\end{equation}
Thus, for most conditions of interest $\mathcal{Y}>>1$, the carbon is fully ionized at the cloud surfaces, and shielding is required to allow the transition from C$^+$ to C.

The integration variables $N_{\rm C^+}$ and $N_{\rm C}$ may be separated by writing
\begin{equation}
 e^{+( \sigma_{\rm g,\odot}/A ) (N_{\rm C^+}/\langle\mu\rangle) } d{N}_{\rm C^+} = 
 \frac{1}{2}\mathcal{Y} \displaystyle\int\displaylimits_0^{1} e^{-( \sigma_{\rm g,\odot}/A ) (N_{\rm C}/\mu)} d\mu \, d{N}_{\rm C} \\\ .
 \label{eq:NCp_NC_separation}
\end{equation}
To enable this separation we have approximated $\mu$ by a constant average $\langle\mu\rangle$ on the left-hand side.
On the right hand side we keep $\mu$ as a variable in the angular integration.
This approximation is similar to that used in \citet{Sternberg2014} to develop Eq.~(\ref{eq:N_HI_total}) for the \HI\ column.
When comparing to numerical calculations (\S~\ref{sec:numerical_models}), we find that setting $\langle\mu\rangle\approx1$ provides a good fit to the results of the numerical radiative transfer computations.

Equation~(\ref{eq:NCp_NC_separation}) is a functional relationship, $N_{\rm C^+}(N_{\rm C})$, between the C$^+$ column and the C column.
Because the C$^+$ column is set by the depth dependent balance between neutralization and ionization processes involving just C and C$^+$, without reference to CO (the transition to C occurs before the transition to CO), we may develop an expression for the total C$^+$ column by integrating
\begin{equation}
 \displaystyle\int\displaylimits_0^{N_{\rm C^+}^{\rm tot}} e^{+( \sigma_{\rm g,\odot}N_{\rm C^+} ) / (A\langle\mu\rangle) } d{N}_{\rm C^+} = 
 \frac{1}{2}\mathcal{Y} \displaystyle\int\displaylimits_0^{\infty} \displaystyle\int\displaylimits_0^{1} e^{-( \sigma_{\rm g,\odot}N_{\rm C} ) / (A\mu)} d\mu \, d{N}_{\rm C} \\\ ,
 \label{eq:C+_integration}
\end{equation}
with $N_{\rm C} \to \infty$ on the right hand side, and $N_{\rm C^+} \to N_{\rm C^+}^{\rm tot}$ on the left.
This gives
\begin{equation}
 N_{\rm C^+}^{\rm tot} = \frac{A\langle\mu\rangle}{\sigma_{\rm g,\odot}} \ln\left( \frac{1}{4}\frac{\mathcal{Y}}{\langle\mu\rangle} + 1 \right) \\\ .
 %\label{eq:N_C+_dust_only}
\end{equation}
The mean $\langle\mu\rangle$ that appears in this expression may now be dropped since our numerical results show that setting $\langle\mu\rangle \approx 1$ provides the best fit.
One may expect $\langle\mu\rangle \approx 1$, since the C$^+$ layer extends to a few optical depths, and photons traveling at even moderate angles from the normal direction are already absorbed at much lower depths.
Our basic expression for the C$^+$ column density is then
\begin{equation}
 N_{\rm C^+}^{\rm tot} = \frac{A}{\sigma_{\rm g,\odot}} \ln\left( \frac{1}{4}\mathcal{Y} + 1 \right) \\\ .
 \label{eq:N_C+_dust_only}
\end{equation}
Importantly, the characteristic column $A/\sigma_{\rm g,\odot} = 7.5\times10^{16}$~cm$^{-2}$ is a constant independent of $Z^\prime$.
For carbon and dust abundances both scaling with the metallicity, the C$^+$ layer extends to a greater linear depth as $Z^\prime$ is reduced, but the greater extent is compensated by the reduced carbon abundance, giving rise to the characteristic column.

\begin{figure}[t]
 \centering
 \includegraphics[width=\columnwidth]{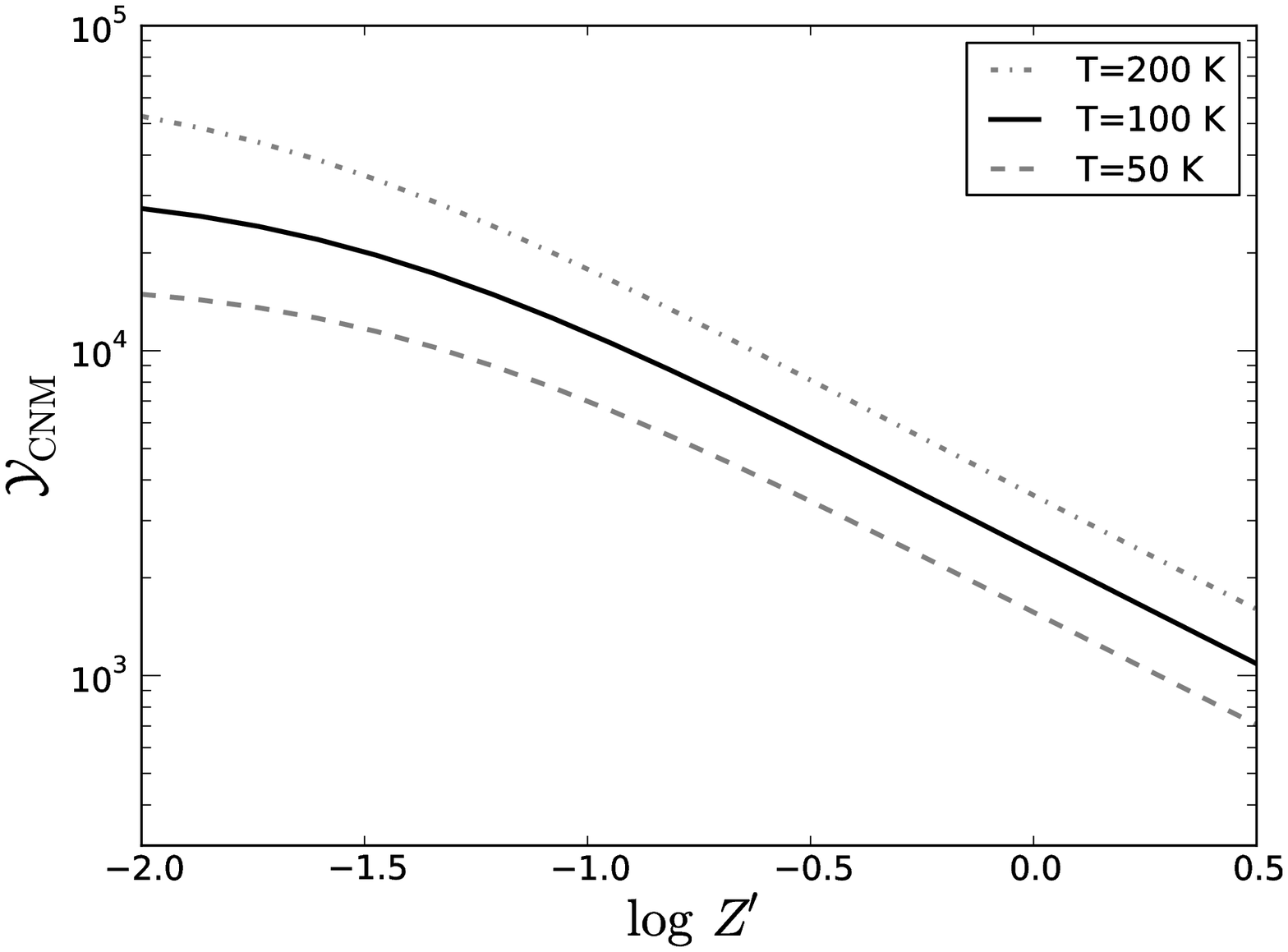}
 \caption{ $\mathcal{Y}_{\rm CNM}$ as a function of $Z^\prime$ at various temperatures.}
 \label{fg:Y_CNM_vs_log_Z}
\end{figure}

For self-regulated gas in which $n\approx n_{\rm CNM}$ as given by Eq.~\ref{eq:CNM_relation}
\begin{eqnarray}
 \mathcal{Y} &=& \mathcal{Y}_{\rm CNM} \equiv \phi_{\rm H_2} \frac{\Gamma}{R_{\rm eff}} \frac{1+3.1Z^{\prime 0.365}}{93} = \nonumber \\
  &=& 2.4\times10^3 \phi_{\rm H_2} \left( \frac{1 + 3.1 Z^{\prime0.365}}{4.1} \right) \left( \frac{\rm 5.5\times10^{-15}\,cm^3\,s^{-1}}{R_{\rm eff}} \right) \,.
 \label{eq:Y_CNM}
\end{eqnarray}
Here, $\phi_{\rm H_2} \lesssim 1$ is a gas ``compression factor'' that allows for the possibility that the H$_2$ component mixed with C$^+$ is denser than any multiphased \HI\ in an outer atomic layer. 
In Figure~\ref{fg:Y_CNM_vs_log_Z} we plot $\mathcal{Y}_{\rm CNM}$ as a function of $Z^\prime$ ranging from $10^{-2}$ to 3, for $T=50$, 100, and 200~K.
For this range of metallicities and temperatures $\mathcal{Y}_{\rm CNM}$ varies from $7\times10^2$ to $5\times10^4$.
For T=100~K, and $Z^\prime=1$, $\mathcal{Y}_{\rm CNM}=2.4\times10^3$, and then as given by Eq.~\ref{eq:N_C+_dust_only} $N_{\rm C^+}^{\rm tot} = 4.8 \times 10^{17}$ cm$^{-2}$.
Because $\mathcal{Y}_{\rm CNM}>>1$, the C$^+$ column varies only logarithmically and is insensitive to the value of $\mathcal{Y}$ as set by $I_{\rm UV}/n$ or $Z^\prime$.

Given our expression for the C$^+$ column density, the total {\it hydrogen} gas column {associated} with the C$^+$ column is,
\begin{equation}
 N_{\rm C^+/(HI+H_2)}^{\rm tot} \equiv
 \frac{N_{\rm C^+}^{\rm tot}}{AZ^\prime}
 = \frac{1}{\sigma_{\rm g}} \ln\left( \frac{1}{4} \mathcal{Y} + 1 \right) \\\ .
 \label{eq:N_C+_dust_only_total_H_HI_incl}
\end{equation}
This includes all of the \HI\ and H$_2$ mixed with the C$^+$.
Equation~(\ref{eq:N_C+_dust_only_total_H_HI_incl}) is very similar to Eq.~(\ref{eq:N_HI_total}) for the \HI\ column density.
The ``${\rm C^+/(\HI+H_2)}$'' column is equal to $1/\sigma_{\rm g}$ multiplied by a logarithmic factor.
However, because $\mathcal{Y} >> \alpha G$, and unless $\alpha G$ is unusually large, the \HI\ column (as given by Eq.~\ref{eq:N_HI_total}) is only a small fraction of the ${\rm C^+/(\HI+H_2)}$ column.
For example, for $Z^\prime=1$ and for $\alpha G = \alpha G_{\rm CNM} = 2.5$ and $\mathcal{Y} = \mathcal{Y}_{\rm CNM}=2.4\times10^3$,
we get $N_{\rm HI}=2.6\times10^{20}$ cm$^{-2}$, but $N_{\rm (HI+H_2), C^+}^{\rm tot}=3.6\times10^{21}$ cm$^{-2}$.
In general, the \HI\ column is less than 10\% of the H$_2$ hydrogen column mixed with the C$^+$, for $\alpha G \lesssim 4.5$.
Thus, for $I_{\rm UV}/n \lesssim 0.1$ cm$^3$ the hydrogen column associated with the C$^+$ may be assumed to be fully molecular.
This \CpHt\ hydrogen column density is then given by
\begin{equation}
 N_{\rm C^+/H_2} \approx
  \frac{1}{Z^\prime \sigma_{\rm g,\odot}} \ln\left( \frac{1}{4} \mathcal{Y} + 1 \right) \\\ .
 \label{eq:N_C+_dust_only_total_H}
\end{equation}
Neglecting the weak dependence on metallicity of the logarithmic factor, the \CpHt\ column scales inversely with $Z^\prime$.

The total LW-band optical depth associated with C$^+$ layer is: $\tau_{\rm LW} = N_{\rm C^+/H_2} \cdot (\sigma_{\rm g,\odot} Z^\prime)$.
Therefore, as with $N_{\rm C^+}^{\rm tot}$ above, $\tau_{\rm LW}$ is almost invariant.
Assuming the CNM relation, the effective optical depth of the C$^+$ layer is nearly constant at $\tau_{\rm LW} \approx 6.5$, or equivalent visual extinction of $A_V\approx2$.

%----------------------------------------------------
\subsection{Inclusion of H$_2$-line blocking}
\label{sec:H2_line_blocking}
%---------------------------------------------------

In addition to dust, H$_2$ also absorbs carbon ionizing photons in discrete LW-band transitions.
The narrow Doppler cores of the H$_2$ lines become optically thick at low H$_2$ columns, but radiation between the lines remains available to photoionize the carbon.
At sufficiently large H$_2$ columns the damping wings of the absorption lines overlap and shield the carbon, in addition to the dust shielding \citep{deJong1980, Tielens1985a}.
This effect becomes especially important at low-$Z^\prime$, where the \CpHt\ hydrogen column (as given by Eq.~(\ref{eq:N_C+_dust_only_total_H})) becomes large.

As we discuss in Appendix~\ref{app:NCp_1st_order_correction}, the (complicated) attenuation of the carbon-ionizing radiation due to H$_2$ line-blocking may be approximated as an exponential attenuation.
Along rays in the normal direction the specific intensity then decreases as,
\begin{equation}
 I_{\rm UV} = I_{\rm UV}(0) e^{-\tau_{\rm dust}} e^{-\tau_{\rm H_2}} \\\ ,
 \label{eq:I_UV with tau_H2}
\end{equation}
where as before $\tau_{\rm dust} \equiv \sigma_{\rm g} N = \sigma_{\rm g} (N_{\rm HI}+2N_{\rm H_2})$, and 
\begin{equation}
 \tau_{\rm H_2} \equiv 2\sigma_{\rm H_2}N_{\rm H_2} \\\ .
\end{equation}
Here, $\sigma_{\rm H_2}$ is an effective cross-section that we evaluate by fitting to the results of numerical computations
for the dependent radiation field including the effects of
varying metallicity
(see Appendix~\ref{app:NCp_1st_order_correction}).
To eliminate factors of two in our expressions below, we define $\sigma_{\rm H_2}$ as the effective H$_2$ line-blocking opacity per hydrogen nucleon, by analogy to $\sigma_{\rm g}$ which is the dust opacity per hydrogen nucleon.
We find that for $Z^\prime \gtrsim 10^{-2}$ and for $\log(I_{\rm UV}/n)>-4$, setting 
\begin{equation}
 \sigma_{\rm H_2} = 7.5 \times 10^{-23} \quad {\rm cm^{-2}}
\end{equation}
provides a good representation for the H$_2$ line blocking opacity.
Importantly, $\sigma_{\rm H_2}$ is independent of $Z^\prime$ and line-blocking becomes increasingly important compared to the dust opacity as the metallicity becomes small.
Our Equation \ref{eq:I_UV with tau_H2} for the attenuated radiation is somewhat simpler than the expression given by \citet{deJong1980} and \citet{Tielens1985a},
who also split the attenuation into dust and H$_2$-line absorption components,
and also include a term for CI continuum absorption.
As we discuss in \S~\ref{sec:neutral_C}, CI absorption is generally negligible however \citep[see also][]{Sternberg2014}.

When neglecting the \HI\ column, $N=2N_{\rm H_2}$, and the total optical depth due to dust plus line-blocking is then
\begin{equation}
 \tau = (\sigma_{\rm g,\odot} Z^\prime + \sigma_{\rm H_2})N = \frac{\sigma_{\rm g,\odot}+(\sigma_{\rm H_2}/Z^\prime)}{A} (N_{\rm C}+N_{\rm C^+}) \\\ .
\end{equation}
Following the identical steps leading to Eq.~(\ref{eq:N_C+_dust_only}) we then have a corrected expression for the C$^+$ column density
\begin{equation}
 {N}_{\rm C^+}^{\rm tot} = \frac{A}{\sigma_{\rm g,\odot}+(\sigma_{\rm H2}/Z^{\prime})} \ln\left( \frac{1}{4}\mathcal{Y} + 1 \right) \\\ ,
 \label{eq:N_C+_tot}
\end{equation}
that includes the effects of H$_2$-line blocking.
The corresponding \CpHt\ hydrogen column density is then
\begin{equation}
 N_{\rm C^+/H_2} = \frac{1}{Z^\prime\sigma_{\rm g,\odot} + \sigma_{\rm H2}} \ln\left( \frac{1}{4}\mathcal{Y} + 1 \right) \\\ .
 \label{eq:N_C+_total_H}
\end{equation}
For $\mathcal{Y} = \mathcal{Y}_{\rm CNM}$, this \CpHt\ hydrogen column is
\begin{equation}
 N_{\rm C^+/H_2}(Z^\prime) = \frac{1}{\sigma_{\rm g,\odot}Z^\prime + \sigma_{\rm H2}} \ln\left( \frac{1}{4}\mathcal{Y_{\rm CNM}} + 1 \right) \\\ ,
 \label{eq:N_C+_tot_H_CNM}
\end{equation}
and is a function of the metallicity only.
For $Z^\prime \lesssim 0.1$, $\sigma_{\rm H_2}$ is larger than $Z^\prime \sigma_{\rm g,\odot}$ and most of the shielding is provided by H$_2$-line blocking.
When $Z^\prime > 0.1$, $\sigma_{\rm H_2}$ is negligible and we recover our ``dust-only'' expressions (\ref{eq:N_C+_dust_only}) and (\ref{eq:N_C+_dust_only_total_H}).

%%%%%%%%%%%%%%%%%%%%%%%%%%%%%%%%%%%%%%%%%%%%%%
%
\section{C$^+$/C/CO Transitions for Varying Metallicity} \label{sec:numerical_models}
%
%%%%%%%%%%%%%%%%%%%%%%%%%%%%%%%%%%%%%%%%%%%%%%

\subsection{Numerical Models}

In order to be able to test the analytic calculation we use a numerical chemistry and radiation transfer code. The MEUDON PDR code\footnote{\url{http://pdr.obspm.fr/PDRcode.html}} \citep{LePetit2006} uses full line-resolved radiative transfer and a large chemical reactions library to solve for the steady state abundance structures of the cloud.
One of the key features of the code is that it allows us to choose an isotropic radiation field as the illuminating source.

A grid of models is created in which we vary the radiation field ($1 \leq I_{\rm UV} \leq 10^3$), density ($10^2 \leq n \leq 10^4$~cm$^{-3}$) and metallicity ($0.01\leq Z^\prime \leq 3.0$).
Inside this parameter space we also calculate several series of models that adopt the CNM relation between $I_{\rm UV}$, $n$ and $Z^\prime$.
The gas-to-dust ratio and $N_{\rm H}$/$E(B-V)$ ratio (`gratio' and `cdunit' parameters in the PDR code) are linearly coupled to the metallicity.
Cosmic radiation is initially turned off.
The latter has little effect on the $\mathrm{C}^+ \rightarrow \mathrm{C}^0$ transition at the shell, but has a more significant effects on the C$\to$CO transition and processes inside the CO core.
We also run the same models with CR turned on at a few times galactic level (H$_2$ ionization rate of $10^{-16}$~s$^{-1}$) to verify that the effect on $N_{\rm C^+}^{\rm tot}$ is not very significant.
Each model is set up as a constant temperature (100~K), constant density, semi-infinite slab and integrated till a depth at which A$_V$=12 and C$^+$/C/CO abundances are no longer a function of depth.
Such a depth is well into the CO core and into regions where in real clouds the temperature significantly decrease and the density increase.
However, we are interested only in the results until the transition to CO ($A_{\rm V} \approx $2--3, at the edge of the temperature(density) drop(rise), and therefore the uniform temperature and density assumptions are acceptable.

\begin{figure*}[p]
 \centering
 \includegraphics[width=\textwidth]{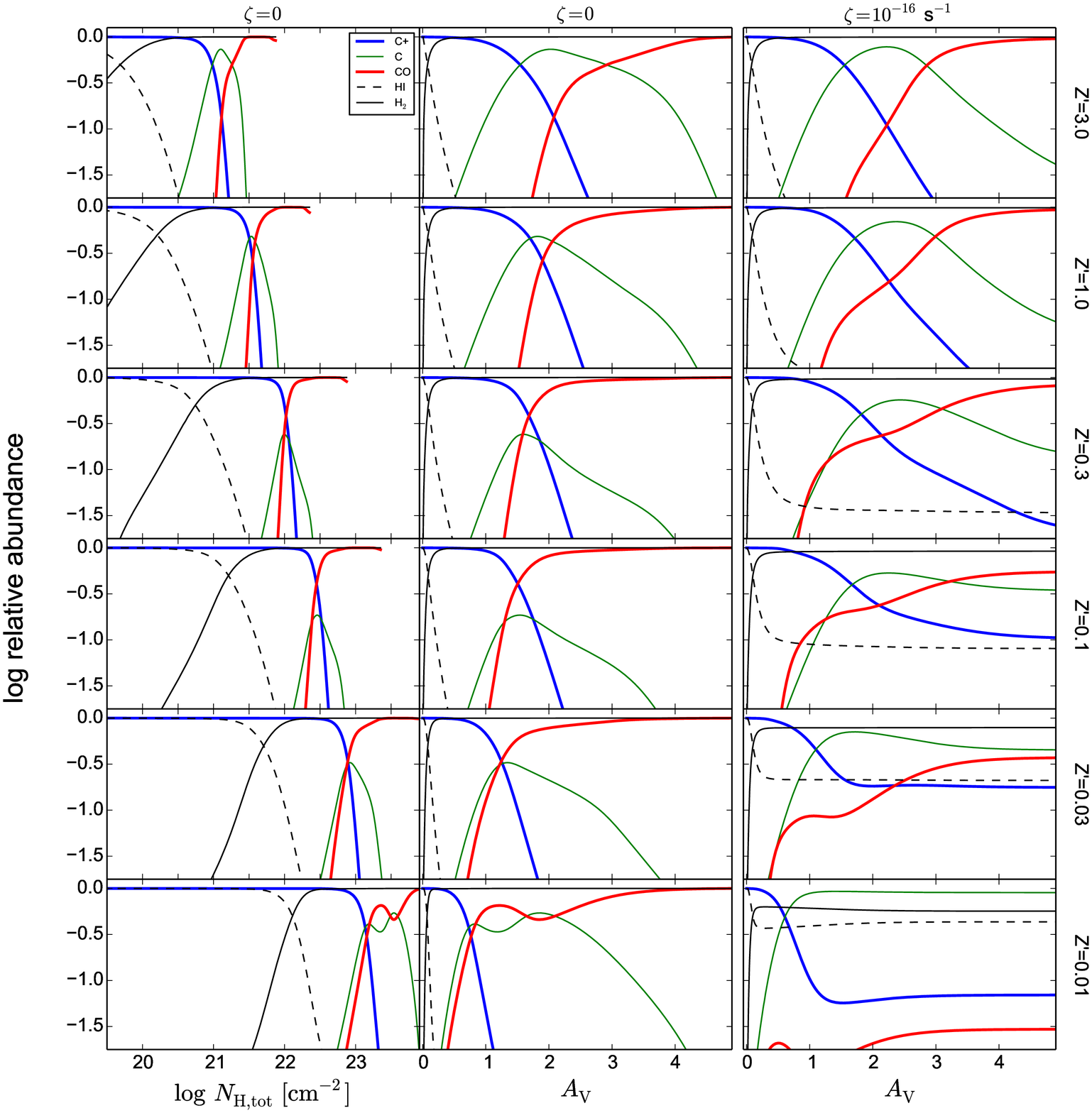}
 \caption{The abundance structure of the molecular cloud for $I_{\rm UV}=10$ and densities assuming the CNM relation ($n\sim400$~cm$^{-3}$).
 The metallicity for each row is indicated on the right and decrease from top to bottom.
 \HI\ and H$_2$ abundances are normalized to total hydrogen abundance. C$^+$, C and CO abundances are normalized to the total carbon abundance.
 {Left-hand column:} as a function of total hydrogen column $N_{\rm H,tot}$ and no CR.
 {Middle column:} same models as on the left-hand column, but plotted as a function of $A_{\rm V}$.
 {\it Right-hand column:} models using similar conditions as the ones in the middle column, except with CR turned on at H$_2$ ionization rate of $\zeta=10^{-16}$~s$^{-1}$.
 %The vertical green and red dotted lines indicate the depth from which C ionization and CO destruction (respectively) are dominated by CR.
 }
 \label{fg:abundance_structure_combined}
\end{figure*}

The abundance structures as calculated by the PDR code for several models assuming CNM conditions are plotted in Figure~\ref{fg:abundance_structure_combined}.
Abundances are normalized to total H in the cases of \HI/H$_2$, and total C in the cases of C$^+$/C$^0$/CO.
The left-hand and center columns plot the same models as a function of total gas column $N$ and $A_{\rm V}$ respectively.
In these models the CR are turned off.
The right-hand column plots models with the same conditions as the center and left-hand columns except for the CR which are set to H$_2$ ionization rate of $\zeta=10^{-16}$~s$^{-1}$ (\S~\ref{sec:CR}).
In each column the metallicity decrease from $Z^\prime=3.0$ at the top panel to $Z^\prime=0.01$ at the bottom panel, as indicated to the right of each row of panels.

%-----------------------------------------------
\subsection{The effect of cosmic rays}
\label{sec:CR}
%----------------------------------------------

In the numerical models described in Section~\ref{sec:numerical_models}, as well as in our analytic calculation of $N_{\rm C^+}^{\rm tot}$ we have ignored the effects of CR.
In CR, we include energetic particle and hard ionizing photons (X-rays) together, since both penetrate deep into the cloud and have a similar effect on the chemistry.
CR are very important in molecular clouds since they penetrate deep into the core of the clouds and create ionization in regions completely shielded from UV.
However, in the outer regions, where the LW band is not completely blocked, UV photons tend to dominate the ionization.
Hence, with respect to the transition from C$^+$ to C, we do not expect CR processes to be very significant, but deeper into the cloud and in particular the transition from C to CO can be sensitive to CR.

We have recalculated the numerical models described in Section~\ref{sec:numerical_models} with the addition of CR, setting the H$_2$ ionization rate to $\zeta=10^{-16}$~s$^{-1}$.
This ionization rate is slightly higher than the local rate in Milky Way.
Because CR have a low cross section for interaction, their intensity tends to represent the mean galactic activity.
UV on the other hand does not propagate far in the galactic disc and UV intensity tends to represent local activity of star formation.
Young stars are typically found in the immediate vicinity of molecular clouds and thus the UV to CR intensity ratio on the edges of molecular clouds tends to be above the galactic mean.
In the following discussion we will focus on the results for the models assuming $I_{\rm UV}=10$, i.e. UV/CR intensities ratios are nearly 10 times higher than the local Milky Way values. The conclusions however are not very sensitive to the exact values.

\begin{figure}[t]
 \centering
 \includegraphics[width=\columnwidth]{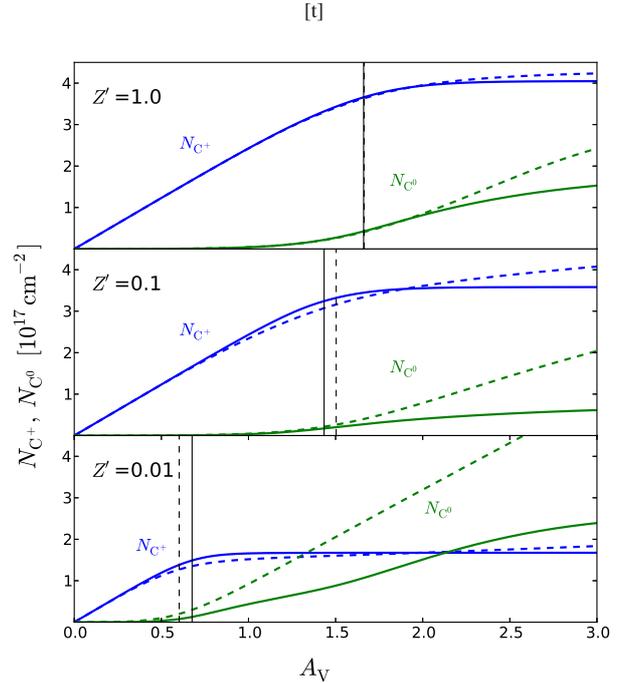}
 \caption{$N_{\rm C^+}$ (blue) and $N_{\rm C^0}$ (green) versus $A_{\rm V}$ for models assuming the CNM relation. Metallicity decreases from top to bottom panels. Models without CR are plotted in solid and models with CR turned on ($\zeta=10^{-16}$~s$^{-1}$) are dashed. The vertical lines indicate the $A_{\rm V}$ at which local C$^+$ abundance is half of the total carbon (dashed/solid indicate with/out CR).}
 \label{fg:N_Cp_vs_Av}
\end{figure}

In the right-hand column of Figure~\ref{fg:abundance_structure_combined} we plot the abundance structure for the same conditions as in the left-hand and middle columns of the figure, except that CR are turned on.
When comparing to the middle column, it is easy to see that the depth at which C$^+$ is no longer the dominant form of carbon hardly changes when CR are turned on.
This is more quantitatively illustrated in Figure~\ref{fg:N_Cp_vs_Av} where we plot the cumulative $N_{\rm C^+}$ column as a function of $A_{\rm V}$ with and without CR.
The depth of the \CpHt\ layer is quite insensitive to the presence of CR.

With CR on, the abundance structures from the C$^+$/C transition and deeper into the cloud become sensitive to the metallicity.
Going deeper into the C and CO dominated regions, the C$^+$ abundance drops more moderately with depth when CR are on.
At the lowest metallicities, a non-zero abundance of C$^+$ is prevalent even in the core of the cloud, and is maintained by CR-induced processes.
The most significant effects are the shift of the transition from C to CO into higher depths than in the no-CR case, and the final balance between C and CO at the core.
Thus, at low $Z^\prime$ a broader region of neutral C exists before the CO core.
When the metallicity is low enough ($Z^\prime<0.1$), neutral C may remain the dominant form of carbon even at depths where there are no UV photons.

Only after all the UV is completely extinguished, including at lower energies than the LW-band that can still dissociate molecules such as OH (important channel in producing CO), does the rate of CO production increase and the CO abundance increase with it till its final CO-core value.
One must keep in mind that in reality the temperatures and densities in the CO core may change significantly relative to the C$^+$/H$_2$ envelope.
Our numerical model assumes uniform temperature and density, and thus the abundance results for the inner CO-core may not be perfectly accurate, though the fundamental effects are preserved.

Since in this paper our interest is the C$^+$ dominated region outside the CO-core, we will ignore CR.
While some abundance of C$^+$ can be maintained by CR-induced processes deep inside the cloud, the expected contribution to the [C II] 158~$\mu$m emission line (which will be discussed later) from these regions is small.
In case the temperature at the core is indeed lower than in the \CpHt\ region, this will make the [C II] 158~$\mu$m emission from the core lower yet.
%This is due to the low temperatures at the core ($\sim$10~K) that are much lower than the equivalent temperature required to excite the [C II] 158~$\mu$m line (91~K), and possible optical depth effects.

%--------------------------------------------------
\subsection{Comparison to Analytic Formulae}
%--------------------------------------------------

\begin{figure}[t]
 \centering
 \includegraphics[width=\columnwidth]{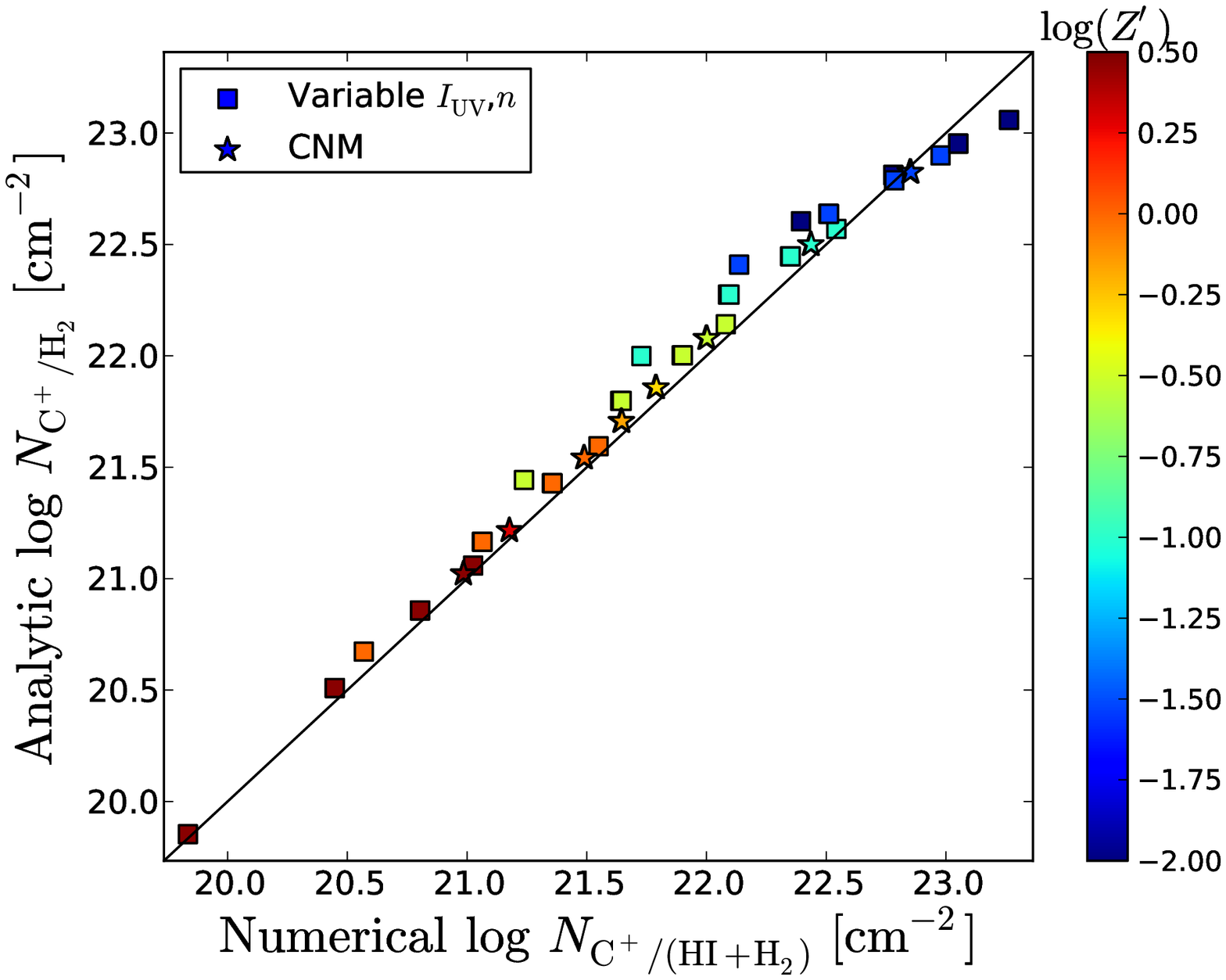}
 \includegraphics[width=\columnwidth]{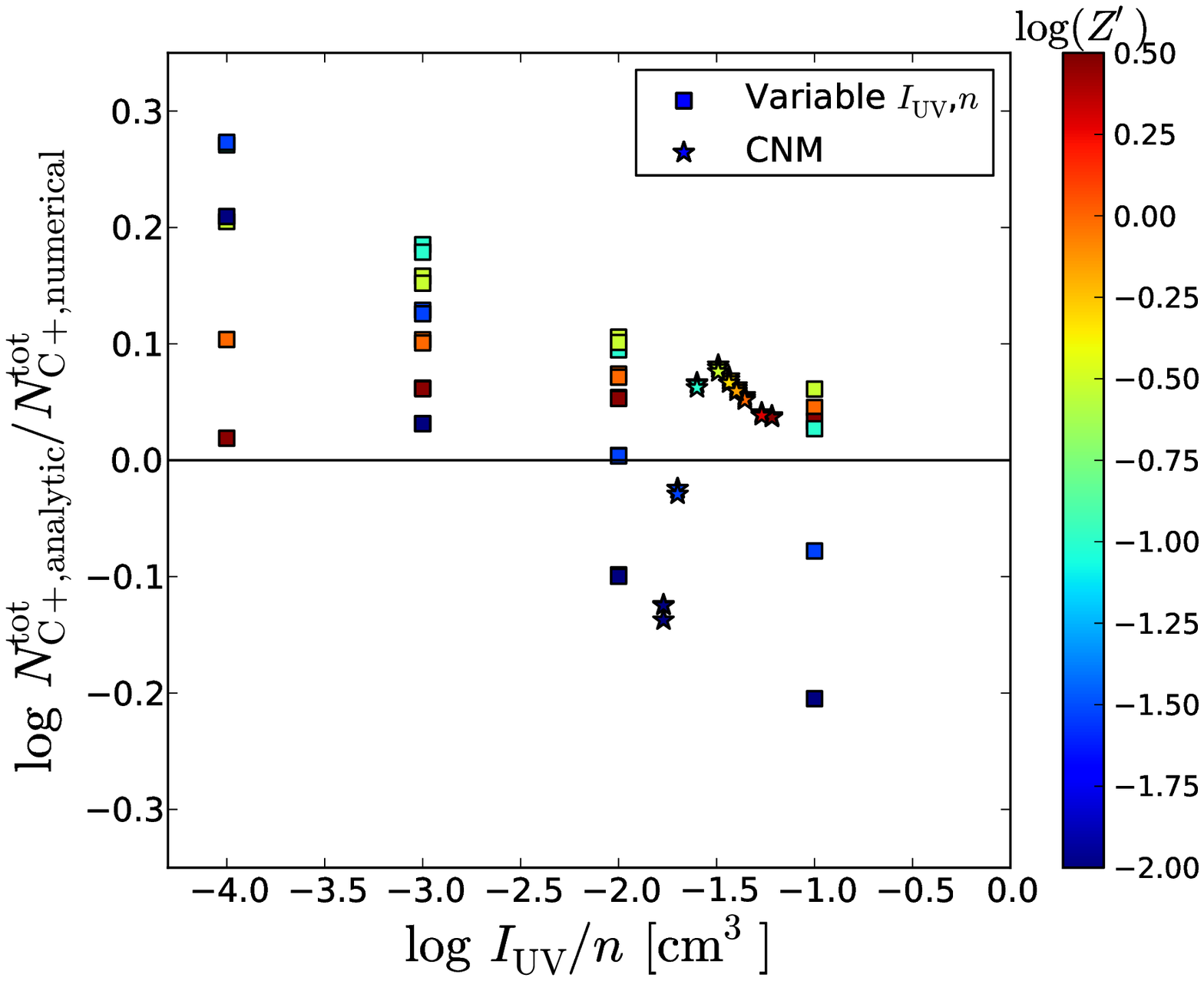}
 \caption{Comparison of the C$^+$ columns as calculated by the analytic expression with the values numerically calculated with PDR code for all models in our models grid.
 {Top}: a comparison of the analytic total gas columns associated with C$^+$ (Eq.~\ref{eq:N_C+_total_H}) with the numerical results.
 {Bottom}: the ratio of the analytic (Eq.~\ref{eq:N_C+_tot}) to numerical total C$^+$ columns as a function of $I_{\rm UV}/n$.
 In both panels, color scale indicates the metallicity.
 }
 \label{fg:NCp_analyt_Meudon}
\end{figure}

In Figure~\ref{fg:NCp_analyt_Meudon} we compare the total \CpHt\ gas column $N_{\rm C^+/H_2}$ predicted by Eq.~\ref{eq:N_C+_total_H} with the column as calculated by the PDR code for our models grid described in \S~\ref{sec:numerical_models}.
The numerically calculated columns include the HI column, which is a mean 6\% of the ${\rm C^+/(\HI+H_2)}$ gas column, and it exceeds 12\% in only one model.
Generally, when restricting to $0.01<z<3$ and $-4<\log(I_{\rm UV}/n)<-0.5$~cm$^3$ the accuracy is within $\pm0.2$~dex, and for models assuming the CNM relation it is within $\pm0.1$~dex.
A turnover for the models with $Z^\prime<0.1$ is clearly visible in the bottom panel and caused by the change from dust shielding to H$_2$ shielding of C.
For $Z^\prime \sim 0.01$ H$_2$ shielding completely dominates and Eq.~\ref{eq:N_C+_tot} becomes very sensitive to the fixed $\sigma_{\rm H_2}$ that we selected.

\begin{figure}[t]
 \centering
 \includegraphics[width=\columnwidth]{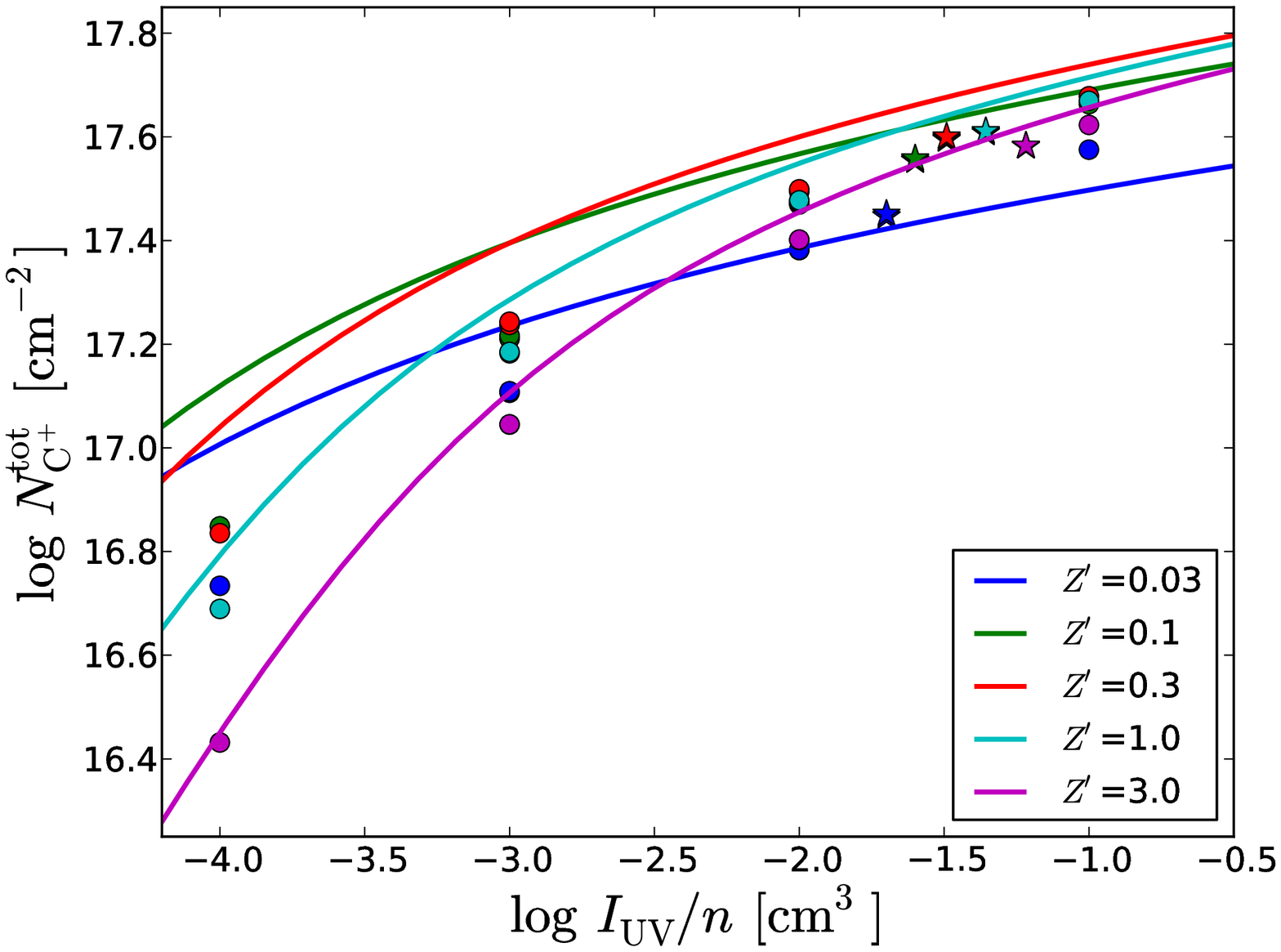}\\
 \includegraphics[width=\columnwidth]{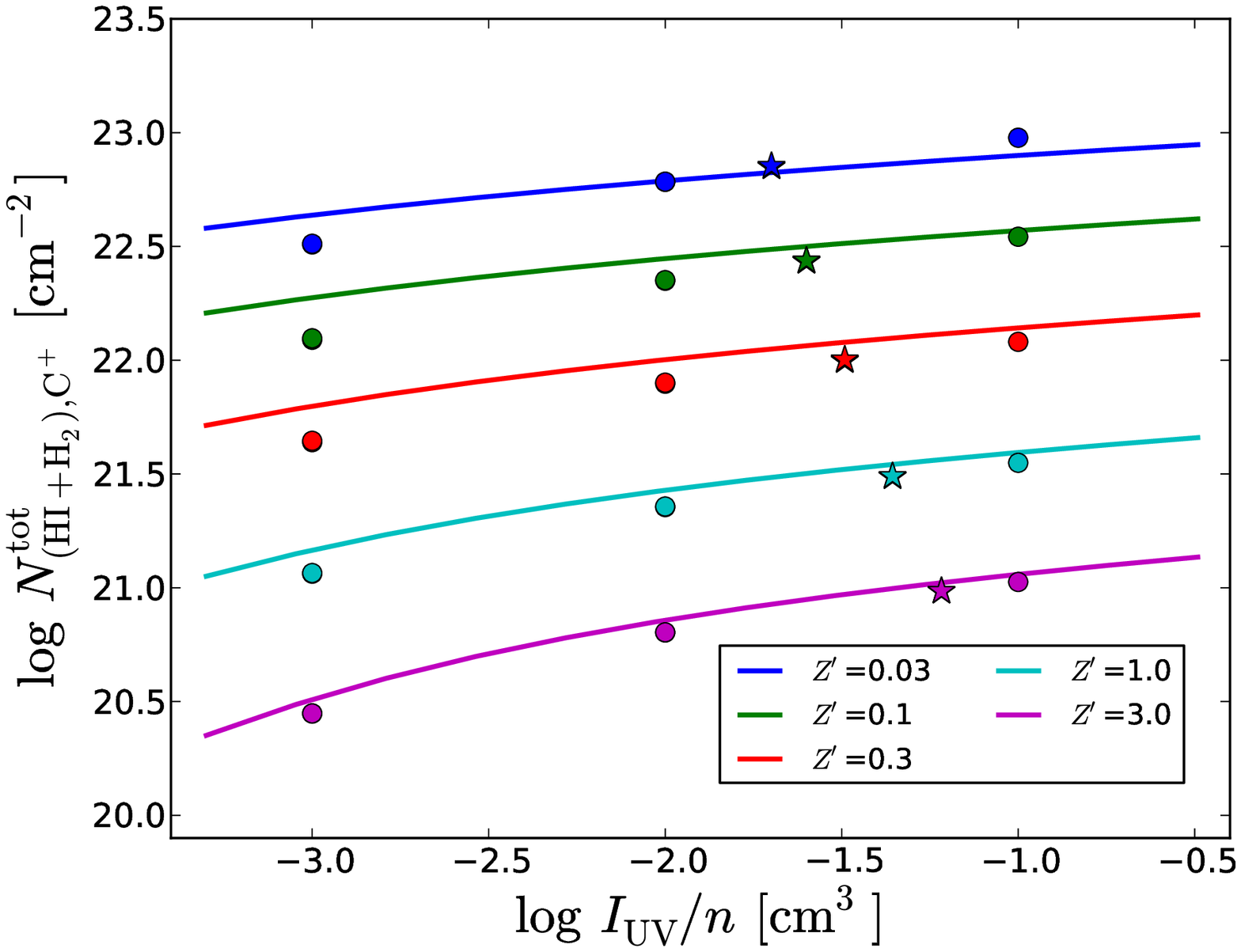}
 % NCp_vs_I_over_n.eps: 0x0 pixel, 300dpi, 0.00x0.00 cm, bb=13 175 598 616
 \caption{{Top:} ${N}_{\rm C^+}^{\rm tot}$ as a function of $I_{\rm UV}/n$ ratio, for various metallicities.
 Star symbols indicate the results of numerical models assuming CNM relation between $I_{\rm UV}$ and $n$.
 Circles indicate numerical models grid with independent $I_{\rm UV}$ and $n$.
 The relation obtained from the analytic calculation (Eq.~\ref{eq:N_C+_tot}) is plotted as solid lines.
 {Bottom:} similar to the above except the vertical axis is the total gas column and the solid lines plot Eq.~\ref{eq:N_C+_total_H}.
 }
 \label{fg:Ncp_vs_I_over_n}
\end{figure}

In Figure~\ref{fg:Ncp_vs_I_over_n} top panel we plot $N_{\rm C^+}^{\rm tot}$ versus $I_{\rm UV}/n$.
The solid curves are calculated with Eq.~\ref{eq:N_C+_tot} and data points are the results of the PDR code.
Note the limited range of $N_{\rm C^+}^{\rm tot}$ columns that spans barely one order of magnitude, while the input values $I_{\rm UV}/n$ span three orders of magnitude and $Z^\prime$ spans over two orders of magnitude.
When comparing the shape of the analytic curves with the numerical calculations of the PDR code we find that the match is good for solar metallicity but gets less accurate as metallicity decrease.
This behavior is also reflected in the overall sloped trend seen in Figure~\ref{fg:NCp_analyt_Meudon} bottom panel.
This inaccuracy was introduced by the approximation of H$_2$ shielding as a constant $\sigma_{\rm H_2}$.
In Appendix~\ref{app:NCp_1st_order_correction} we present a more accurate treatment of a variable $\sigma_{\rm H_2}$ that is a function of the column.
With a $\sigma_{\rm H_2}(N)$, the accuracy improves and all trends removed, at the cost of a more complex expression for $N_{\rm C^+}^{\rm tot}$.
However, we feel that an accuracy better than 0.2~dex is often not required, since it may exceeds the accuracy of other modeling assumptions, or observational measurement errors.

%%%%%%%%%%%%%%%%%%%%%%%%%%%%%%%%%%%%%%%%%%%%
%
\section{Neutral C}
\label{sec:neutral_C}
%
%%%%%%%%%%%%%%%%%%%%%%%%%%%%%%%%%%%%%%%%%%%%

In the analytic model we have neglected the UV absorption by neutral C, which requires a justification.
The cross-section for C photoionization is $\sigma_{\rm C}=1.5\times10^{-17}$ cm$^2$.
This means that gas columns of $N \gtrsim 1/(\sigma_{\rm C} A \xi)$ are required for neutral C to shield itself, where $\xi$ is the mean carbon fraction of C$^0$ in the C$^+$ region.
$\xi$ can be estimated from Eq.~\ref{eq:rate_equation} for the case $\tau<<1$.
We get $\xi \approx 6\times10^{-5} (n/I_{\rm UV})$, meaning that the gas column required for neutral C self-shielding is $N \approx 8\times10^{24} (I_{\rm UV}/n)$~cm$^{-2}$.
For the practical cases of $I_{\rm UV}/n > 10^{-4}$~cm$^3$ (and in particular $I_{\rm UV}/n \approx 5\times10^{-2}$~cm$^3$ for CNM) this column is larger than the column for dust shielding $1/\sigma_{\rm g,\odot}=6\times10^{20}$~cm$^{-2}$.

The same conclusion regarding the insignificance of neutral C self-shielding can be reached by looking at the cumulative columns calculated by the PDR code and plotted in Figure~\ref{fg:N_Cp_vs_Av}.
We can see that at the depth at which C$^+$ is no longer the dominant carbon form (indicated by vertical lines) and $N_{\rm C^+}^{\rm tot}$ is almost reached, the neutral C columns of $N_{\rm C^0}(A_{\rm V})<1\times10^{17}$~cm$^{-2}$ are insufficient to provide effective shielding.

In Figure~\ref{fg:abundance_structure_combined} we see that some of the C$^0$ column is mixed with the C$^+$ gas outside the CO region and is part of the `CO-dark gas'.
Unlike ${\rm C}^+ \leftrightarrow {\rm C}^0$ processes, ${\rm C}^0 \leftrightarrow {\rm CO}$ processes are more complicated and a full analytic calculation is outside the scope of this paper.
However, we will argue that in nearly all relevant conditions the C$^+$ column is larger than the C$^0$ column by factors of a few and we can neglect the contribution of C$^0$ to the CO-dark gas column.

\begin{figure}[t]
 \centering
 \includegraphics[width=\columnwidth, clip=true,trim=0pt 0pt 0pt -10pt]{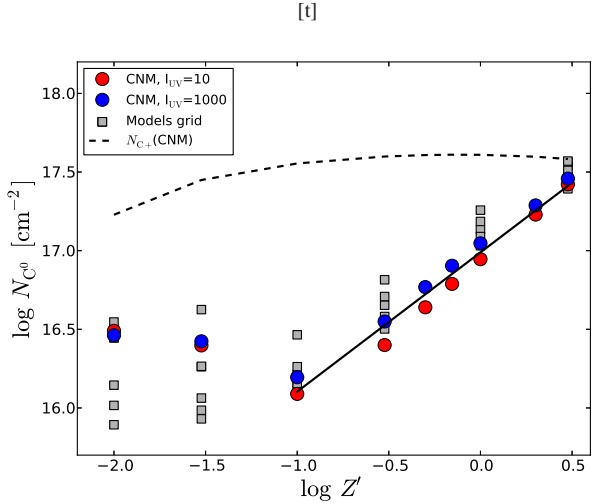}
 \caption{The neutral carbon column outside the CO-dominated region as a function of metallicity, calculated with the PDR code for various $I_{\rm UV}$ and $n$.
 Models that assume CNM relation between $I_{\rm UV}$ and $n$ are plotted as colored circles while the grid of independent $I_{\rm UV}$ and $N_{\rm H}$ models are plotted as gray squares.
 The solid black line is a power-law fitted ad hoc to the CNM points at $\log(Z^\prime)\geq-1$.
 The C$^+$ column assuming CNM conditions, as calculated by the PDR code is plotted for comparison as a broken black line.}
 \label{fg:N_CI}
\end{figure}

In Figure~\ref{fg:N_CI} we plot the total ${N}_{\rm C^0}$ column calculated by the PDR code in the models grid as a function of the metallicity.
We include the neutral C column only until the depth at which $n_{\rm C^0} = n_{\rm CO}$.
The ${N}_{\rm C^+}^{\rm tot}$ column is also plotted for comparison.
At $\log(Z^\prime)\geq-1$ the scatter between the models is small and it seems that $Z^\prime$ is the only important parameter, nearly independent of $I_{\rm UV}$ and $n$.
At all metallicities up to solar, the C$^+$ column is larger than the C$^0$ column that is outside the CO region by factors of a few.
However, the latter increase almost linearly with $Z^\prime$, until at $Z^\prime=3$ it is about equal to the C$^+$ column.
We find that ${N}_{\rm C^0}(Z^\prime)$ for $\log(Z^\prime)\geq-1$ can be approximated by the power-law
\begin{equation}
 \log\left(\frac{{N}_{\rm C^0}}{\rm cm^{-2}}\right)_{n_{\rm C}>n_{\rm CO}} = 0.88 \log(Z^\prime) + 17.0 \,,
 \label{eq:N_CI}
\end{equation}
that is plotted in Figure~\ref{fg:N_CI} as a solid black line.

At $\log(Z^\prime)<-1$ the scatter between various $I_{\rm UV}$ and $n$ starts to be more significant, yet in all cases ${N}_{\rm C^0} << {N}_{\rm C^+}^{\rm tot}$.
Looking at the abundance profiles in Figure~\ref{fg:abundance_structure_combined} (for the case of $\zeta=0$) we can see that a significant neutral C column does exist at $Z^\prime<0.1$, but the vast majority of it is mixed with the CO instead of being mixed with C$^+$ (low metallicity), or forming a neutral C dominated region (super-solar metallicity).

The existence of a significant neutral C column inside the CO region is due to dust shielding becoming less effective than H$_2$ shielding.
H$_2$ shields the LW-band that photoionize C, and also dissociates CO through lines.
Remarkably, all such CO lines are within the LW-band \citep{vanDishoeck1988}.
However, OH which is important for forming CO through ${\rm C + OH} \rightarrow {\rm CO + H}$, can be dissociated by photons of much lower energy than the LW-band and is not fully shielded by H$_2$.
Thus, the formation of CO through reaction with OH is slowed down and even a low flux of LW-band photons is still sufficient to maintain a significant C$^0$ population by dissociating CO.
The LW band flux at those depths is insufficient to maintain a C$^+$ population, thus the C$^+$ column ends, and we get a region of C$^0$ mixed with CO.
This effect can be seen in the middle column of Figure~\ref{fg:abundance_structure_combined} - At high metallicities C$^+$ abundance, as well as the C$^0$ abundance drop at $A_V \approx 2$.
In contrast, at the low metallicities (bottom panels) C$^+$ abundance drops at a lower $A_V$ due to the contribution of H$_2$ shielding.
However, the C at low metallicities stays high till $A_V \approx 2$ and only at higher depths, with sufficient dust shielding to UV photons that are not blocked by H$_2$, C is converted into CO.

From Figure~\ref{fg:abundance_structure_combined} (left-hand and middle columns) it is clear that when there are no CR and metallicities are very low, most of the neutral C column is mixed with CO in the CO dominated region.
The abundance of neutral C in the CO region beyond the PDR is sensitive to the CR flux, as can be seen in the right-hand column of Figure~\ref{fg:abundance_structure_combined}, and discussed in \S~\ref{sec:CR}.
The column of neutral C inside the CO region is not strictly `CO-dark gas'.
Except in the case of super-solar metallicities, where the neutral C column is both significant and mostly outside the CO dominated region, we can generally neglect the contribution of neutral C to the CO-dark gas column.

%%%%%%%%%%%%%%%%%%%%%%%%%%%%%%%%%%%%%%
%
\section{The C$^+$/H$_2$ gas} \label{sec:C+/H2_gas}
%
%%%%%%%%%%%%%%%%%%%%%%%%%%%%%%%%%%%%%%

% The `CO dark gas' is the gas residing inside the H$_2$ molecular region, but outside the CO core, thus no CO lines are emitted from this region.
% As we argued previously, most of this H$_2$ gas has the carbon in it in C$^+$ form, and we refer to this gas as the C$^+$/H$_2$ gas.
% In most cases, neutral C outside the CO region contributes significantly less than C$^+$ to the CO-dark molecular gas. 
% Also, we find that the C$^+$ column in the HI region is only a small fraction of the total C$^+$ column.
% This means that Eq.~\ref{eq:N_C+_total_H} is a good approximation to $N_{\rm C^+/H_2}$ and to the total CO-dark gas column.
% In addition, if the $I_{\rm UV}/n$ values are close to the CNM relation, we can use Eq.~\ref{eq:N_C+_tot_H_CNM} with $Z^\prime$ as the only parameter for $N_{\rm C^+/H_2}$.

\begin{figure}[t]
 \centering
 \includegraphics[width=\columnwidth]{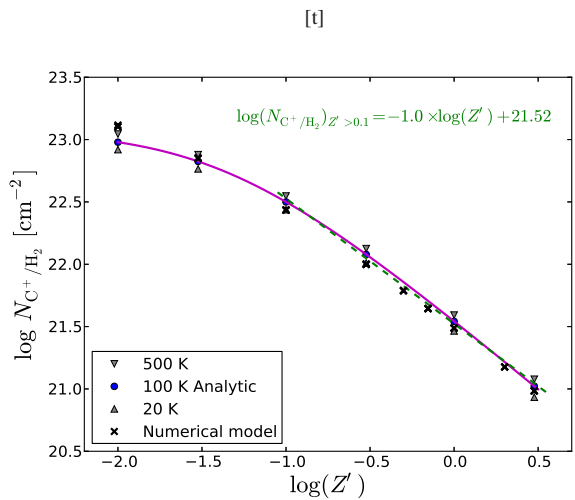}
 \caption{$N_{\rm C^+/H_2}$ as a function of metallicity assuming CNM. The points calculated with Eq.~\ref{eq:N_C+_total_H} and assume $T$=100~K are plotted in blue circles and fitted by a linear relation for $Z^\prime>0.1$ (green line). Variations of the temperatures by factors of 5 (gray triangles) make only a minor difference. Full calculations with the PDR code at 100~K are plotted as black crosses.}
 \label{fg:N_CpH2_vs_Z}
\end{figure}

$N_{\rm C^+/H_2}$ as a function of $I_{\rm UV}/n$ is plotted in Figure~\ref{fg:Ncp_vs_I_over_n} for various metallicities.
Note that the curves are quite flat, especially around $\log(I_{\rm UV}/n) \approx -1.5$ ($n$ in cm$^{-3}$) values typical to CNM conditions, meaning that even large deviations in the $I_{\rm UV}/n$ ratio relative to the CNM ratio have only a small effect on the resulting columns.
Therefore we can assume the relation (Eq.\ref{eq:CNM_relation}) and then $N_{\rm C^+/H_2}(Z^\prime)$ is a function of metallicity only.
This result of $N_{\rm C^+/H_2}(Z^\prime)$ is visualized in Figure~\ref{fg:N_CpH2_vs_Z}, where we plot the `CO-dark' gas column on one side of the CO core
as a function of $Z^\prime$.
We plot both the analytic predictions derived from Eq.~\ref{eq:N_C+_total_H} and the numerical results of the PDR code.
$N_{\rm C^+/H_2}$ is nearly linear with $Z^{\prime-1}$ down to $Z^\prime \sim 0.1$ and very weakly dependent on the temperature, even if we vary it by factors of 5.
At 100~K and $\log(Z^\prime)>-1$ the relation from our analytic formula (Eq.~\ref{eq:N_C+_tot_H_CNM}) is approximated by
\begin{equation}
 \log(N_{\rm C^+/H_2})_{Z^\prime>0.1} \approx -1.0 \log(Z^\prime) + 21.52 \,,
 \label{eq:N_DG(Z)}
\end{equation}
which is fully consistent with the results of the PDR code.

The total amount of CO-dark molecular gas is therefore a function of the metallicity and the total surface area of the cloud, almost regardless of the variations in the radiation fields due to the location in the galaxy or nearby UV sources.
Let us define the \CpHt\ molecular gas fraction (approximately the CO-dark fraction) as:
\begin{equation}
 f_{\rm C^+/H_2} \equiv \frac{2N_{\rm C^+/H_2}}{N_{\rm cloud}}
 \label{eq:f_DG_definition}
\end{equation}
where $N_{\rm cloud}$ is the total column of hydrogen in the cloud from one side to the other.
The factor of 2 account for both sides of the slab as sketched in Figure~\ref{fg:cloud_sketch}.
The numerator, as we demonstrated above depends mostly on the metallicity and very weakly on the details of the UV radiation fields and density.
The implication is that the \CpHt\ gas fraction in a galaxy is a function of the metallicity and the typical size of the clouds in the galaxy.

\begin{figure}[t]
 \centering
 \includegraphics[width=\columnwidth]{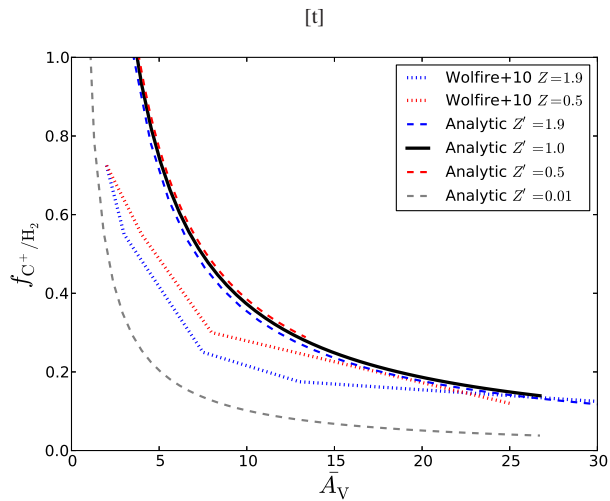}
 % f_DG_vs_Av.eps: 0x0 pixel, 300dpi, 0.00x0.00 cm, bb=13 175 598 616
 \caption{The ${\rm C+/H_2}$ molecular gas fraction as a function of the mean total $A_{\rm V}$ through the cloud from one side to the other.
 The numerical models of W10 (dotted lines) assume a spherical geometry and $\bar{A_{\rm V}}$ refers to the mean optical depth through the cloud. The analytic model assumes CNM conditions.
 Metallicity dependence is weak except when $Z^\prime \lesssim 0.1$, then H$_2$ shielding becomes significant (example, black dashed line for $Z^\prime=0.01$).}
 \label{fg:f_DG_vs_Av}
\end{figure}

In Figure~\ref{fg:f_DG_vs_Av} we plot $f_{\rm C^+/H_2}$ as a function of $\bar{A}_{\rm V}$ the mean total $A_{\rm V}$ from one side of the cloud to the other, using Eq.~\ref{eq:N_C+_tot_H_CNM} for $N_{\rm C^+/H_2}$.
$f_{\rm C^+/H_2}$ reaches 1 at ${\bar A}_{\rm V} \approx 4$, below which the cloud is not thick enough to support a CO core in our model.
Also plotted in Figure~\ref{fg:f_DG_vs_Av} are the results of the numerical model of W10.
Their model assumes a cloud of spherical geometry with a $n\propto r^{-1}$ density profile and a mean projected $\bar{A}_{\rm V}$ value (i.e., averaged over the sphere).
They normalize the density profile so there is always a CO core with a mass of $10^6$~M$_\odot$.
Due to the way their model is constructed, $f_{\rm C^+/H_2}$ can never reach a value of 1 in W10, as opposed to our plane parallel model that requires a minimum column (or $A_{\rm V}^{\rm tot}$) to produce CO.
For this reason the two models diverge at low $A_{\rm V}$.
The models converge when $\bar{A}_{\rm V}$ is high.

From the plot one may get the impression that $f_{\rm C^+/H_2}$ does not depend on $Z^\prime$ directly and it is only the total $A_{\rm V}$ of the cloud that sets $f_{\rm C^+/H_2}$.
This is true only for $Z^\prime \gtrsim 0.1$ where H$_2$ shielding is unimportant and the ${\rm C^+/H_2}$ column scales linearly with $Z^\prime$.
However, at a lower metallicity H$_2$ shielding moves the C$^+ \to $C transition to a lower $A_{\rm V}$ and thus lowers $f_{\rm C^+/H_2}$ for a given $\bar{A}_{\rm V}$.
This effect can be seen in the $Z^\prime=0.01$ curve that we plot in Figure~\ref{fg:f_DG_vs_Av}.

One must keep in mind that in case the total gas column in the cloud is insufficient to support twice the column given by Eq.~\ref{eq:N_C+_total_H}, then this equation can only provide an upper limit to the C$^+$ column. In such cases the cloud will not have a CO core at all, and $N_{\rm C^+}^{\rm tot}$ will be limited by the total gas column of the cloud.
At solar metallicity this implies clouds of $N_{\rm cloud} \gtrsim 6 \times 10^{21}$~cm$^{-2}$ are required in order to support a CO core, and total columns in the range of $10^{23}$~cm$^{-2}$ are required in case of metallicities $Z^\prime \lesssim 0.1$.
However, at such low metallicities CO in the core can be easily destroyed in the presence of CR and result in very large clouds that are non the less almost completely CO-dark.
CO destruction by CR-induced processes is completely different physics/chemistry from the destruction by UV in the PDR region, and each dominates in different locations in the cloud.
When attempting to convert CO line flux measurements to total molecular gas mass the two processes need to be taken into account separately.

%%%%%%%%%%%%%%%%%%%%%%%%%%%%%%%%%%%%%%%
%
\section{The [C II] 158~$\mu\textrm{\lowercase{m}}$ emission} \label{sec:CII_emission}
%
%%%%%%%%%%%%%%%%%%%%%%%%%%%%%%%%%%%%%%%%

The \cii\ line is one of the brightest emission lines observed in star forming galaxies and contributes a significant fraction of the cooling in PDR regions.
In the previous sections we have derived an analytic expression for the column of C$^+$ gas in a molecular cloud.
It is now interesting to investigate the relation between the C$^+$ column and the emerging \cii\ intensity (which we will refer to as simply  [C~{\sevensize II}] for brevity).

%  [C~{\sevensize II}] is excited by collisions with H, H$_2$ and free electrons.
% Since, as we have seen in Section~\ref{sec:analytic_C+_column} most of the C$^+$ ions reside inside the H$_2$ region, the contribution of excitations by H is small.
% Following \citet{Osterbrock1989}, the  [C~{\sevensize II}] cooling rate coefficient due to collisions with free electrons (at T=100~K) is:
% %$\rho_e = 25 T^{-1/2} \exp{-h\nu/kT}$
% $\rho_e = 1.25\times10^{-20}$ erg~s$^{-1}$~cm$^{-3}$.
% Since within $\sim$20\%\ accuracy $n_e \approx n_H X_C$, this gives $\rho_e n_H X_C = 1.65\times^{-24} n Z$~erg~s$^{-1}$ per C$^+$ ion.
% This is in comparison to the H$_2$  [C~{\sevensize II}] cooling excitation rate coefficient: $\rho_{\rm H_2}(T=100K) = 5\times10^{-24}$ erg~s$^{-1}$ \citep{Flower1977}, for which the emission power per ion is $\rho_{\rm H_2} n/2 = 2.5\times10^{-24} n$ erg~s$^{-1}$.
% Therefore, at solar metallicity excitation by electrons is moderately significant, but its contribution decreases rapidly with decreasing metallicity.
% At super-solar metallicities it may become as significant as H$_2$ collisions.

\begin{figure}[t]
 \centering
 \includegraphics[width=\columnwidth]{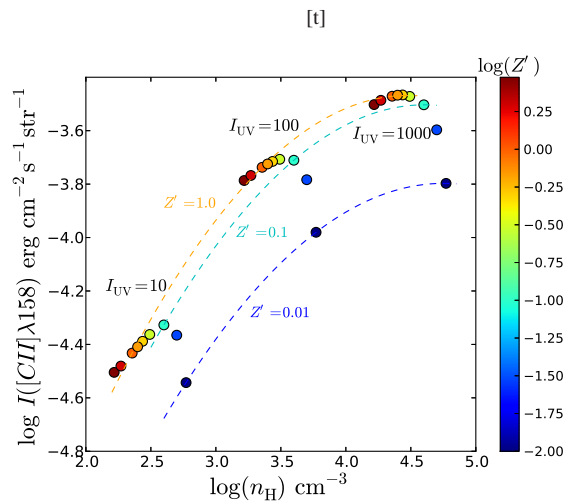}
 % I_CII_vs_nH.eps: 0x0 pixel, 300dpi, 0.00x0.00 cm, bb=13 175 598 616
 \caption{The emerging \cii\ intensity, emitted from the observer side of the slab, as a function of density. Color scale indicates the metallicity. We plot three series of models with different $I_{\rm UV}$ for which the density is set by the CNM relation. Dashed lines are simple 2nd order interpolations at a constant $Z^\prime$.}
 \label{fg:CII_vs_nH}
\end{figure}

The MEUDON PDR code calculates the full radiation transfer of the emitted  [C~{\sevensize II}] photons and returns the observed intensity.
Since we run the code as a semi-infinite slab, the resulting intensity accounts only for photons originating from the side of the slab facing the observer.
 [C~{\sevensize II}] photons from the opposite side are likely to pass through the cloud with little attenuation unless the total C$^+$ column is very large.
In figure~\ref{fg:CII_vs_nH} we plot the  [C~{\sevensize II}] intensity as a function of the total hydrogen density for various radiation fields and metallicities, at 100~K temperature.
In this plot we assume the CNM relation between $I_{\rm UV}$ and $n$, therefore at a given metallicity $I_{\rm UV}$ and $n$ are correlated.
As we can see from the figure, the emerging  [C~{\sevensize II}] intensity follows nearly the same relation for metallicities of $Z^\prime \gtrsim 0.1$, but at lower metallicities the intensity drops significantly.
This is due to H$_2$ shielding of neutral C that lowers the resulting C$^+$ column, while at higher metallicities neutral C is shielded mostly by the dust and the resulting $N_{\rm C^+}^{\rm tot}$ is almost independent of metallicity (Eq.~\ref{eq:N_C+_tot}).

\begin{figure}[t]
 \centering
 \includegraphics[width=\columnwidth]{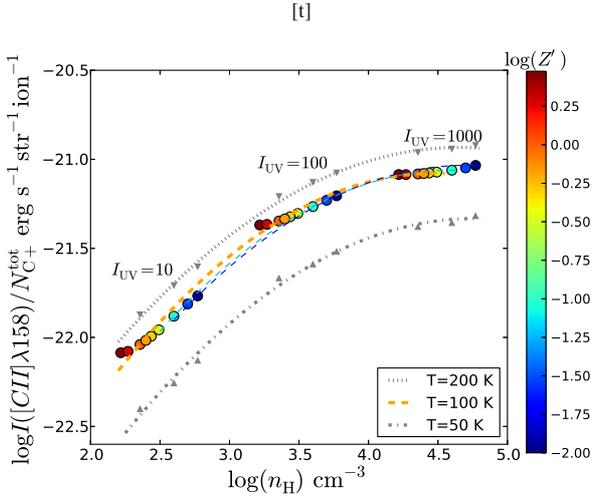}
 % I_CII_over_NCp_vs_nH.eps: 0x0 pixel, 300dpi, 0.00x0.00 cm, bb=13 175 598 616
 \caption{The ratio of \cii\ intensity to the C$^+$ column density. This is the power emitted per C$^+$ ion per solid angle. Color coding, dashed lines, and models plotted are the same as in Figure~\ref{fg:CII_vs_nH}.
 The grey triangular markers and gray lines are for the similar models as the colored markers and lines, but different gas temperatures.}
 \label{fg:CII/N_C+_nH}
\end{figure}

In Figure~\ref{fg:CII/N_C+_nH} we plot the  [C~{\sevensize II}] intensity to $N_{\rm C^+}^{\rm tot}$ ratio versus $n$ as calculated by the PDR code, assuming CNM conditions.
In this plot the meaning of the y-axis is the emitted power per C$^+$ ion.
When plotted in this way, the dependence on metallicity is nearly canceled.
There is a small increase in the  [C~{\sevensize II}] emission per ion over the general trend when $Z^\prime>1$ due to excitations by free-electrons.
The  [C~{\sevensize II}] excitation rate by free electrons \citep{Osterbrock1989} depends linearly on the metallicity, while the excitation rate by H$_2$ collisions \citep{Flower1977} is independent.
At $Z^\prime=1$ electron collisions are moderately significant, but their contribution drop with decreasing $Z^\prime$.
At super-solar metallicities electron collisions may become as significant as H$_2$, which explains the small excess in  [C~{\sevensize II}]/$N_{\rm C^+}^{\rm tot}$ at $Z^\prime=3.0$.

As we can see in the figure, when the radiation fields increase as does the density due to the CNM relation, the emission per ion increase less and less till it tends to a saturation at $\sim 10^{-21}$~erg~s$^{-1}$~str$^{-1}$.
This is due to a critical density of H$_2$ beyond which collisional de-excitation dominates over radiative decay.
Following \citet{Tielens1985a}, the critical density is defined as $n_{\rm crit} \equiv A_{21}/\gamma_{\rm H_2}$, where $A_{21}$ is the Einstein coefficient and $\gamma_{\rm H_2}$ is the collisional de-excitation rate coefficient.
Since most of the  [C~{\sevensize II}] is emitted from the ${\rm C^+/H_2}$ region, the carbon is de-excited by H$_2$ collisions and $n_{\rm crit} \approx 10^4$~cm$^{-3}$ (in total hydrogen nucleons density). 
The  [C~{\sevensize II}] emission per ion saturates at lower densities ($\sim 4\times10^3$~cm$^{-3}$) in the neutral H region, however this represents only a small fraction of $N_{\rm C^+}^{\rm tot}$ (Section~\ref{sec:analytic_C+_column}).

%%%%%%%%%%%%%%%%%%%%%%%%%%%%%%%%%%%%%%
%
\section{Discussion: C$^+$/H$_2$ gas in star forming galaxies} \label{sec:Discussion}
%
%%%%%%%%%%%%%%%%%%%%%%%%%%%%%%%%%%%%%%%

Thanks to modern millimeter interferometers such as the Plateau-de-bure and ALMA, it is now possible to measure molecular gas masses in many galaxies at $z>1$ via observations of CO rotational lines.
Star forming disc-like galaxies account for $\sim$90\% of the total star formation volume density \citep[][]{Rodighiero2011} and in such galaxies very high molecular gas fractions (mean 0.4--0.5) are observed \citep{Daddi2010, Tacconi2010, Tacconi2013}.
Since these gas mass measurements rely on CO lines it is interesting to consider how much molecular gas is unaccounted for due to carbon being in the C$^+$ state - i.e. the fraction of the C$^+$/H$_2$ gas.

Our analytic model predicts the C$^+$/H$_2$ column on the face of the cloud, irrespective of the total cloud size/mass (given it is sufficiently large) and thus, requires more information about the typical clouds sizes if gas fractions are to be calculated.
Under CNM conditions which supposedly described global mean conditions in the galaxy, we have found that the expected $N_{\rm C^+/H_2}$ column is a function of the metallicity only (Eq.~\ref{eq:N_C+_tot_H_CNM} and Figure~\ref{fg:N_CpH2_vs_Z}).
%Thus, we expect a mean {\it two-sided} column of $N_{\rm C^+/H_2} = 6\times10^{21}$~cm$^{-2}$ (equivalent to a mass surface density of $\Sigma=68$~M$_\odot$~pc$^{-2}$) that will scale linearly with $Z^{\prime-1}$ down to $Z^\prime \approx 0.1$.
We can now do the naive exercise of comparing the expected ${\rm C^+/H_2}$ column with the molecular gas columns measured from CO observations in surveys such as the local COLDGASS \citep{Saintonge2011a}, and the $1<z<2.5$ PHIBBS \citep{Tacconi2013}.
For the purpose of our exercise we will assume that the uniform conversion factor used in these surveys to convert from CO luminosity to molecular gas mass accounts only for the molecular gas in the ${\rm CO/H_2}$ phase (i.e., the CO core).

The COLDGASS sample covers the $z=0$ SFR--$M_{\rm star}$ relation for galaxies of M$_{\rm star}>10^{10}$~M$_\odot$.
It is randomly selected and thus represent the typical distribution of galaxy parameters at those masses.
The typical molecular surface densities that they obtain are of the order 20~M$_\odot$~pc$^{-2}$, though a wide range exists within the sample.
Assuming $Z^\prime=1$ in local massive spirals, we expect $N_{\rm C^+/H_2} = 6\times10^{21}$~cm$^{-2}$ {\it two-sided} column that is equivalent to $\Sigma_{\rm C^+/H_2}=68$~M$_\odot$~pc$^{-2}$.
We can define the ${\rm C^+/H_2}$ molecular gas fraction as:
\begin{equation}
 f_{\rm C^+/H_2} = \frac{\Sigma_{\rm C^+/H_2}}{\Sigma_{\rm C^+/H_2} + c\Sigma_{\rm CO/H_2}}
 \label{eq:f_C+_H2}
\end{equation}
where $\Sigma_{\rm CO/H_2}$ is the molecular surface density as derived from CO (galactic conversion factor) and $c$ is a clumpiness factor.
The molecular gas is not evenly spread over the galaxy. Instead, it tends to be concentrated in large clumps.
This clumpiness factor (inverse of the filling factor) represents the larger column densities at the locations of the clouds over the global mean.
\citet{Krumholz2011} for example argues \citep[based on][]{Krumholz2009c} a clumpiness factor $c=5$.
\citet{Leroy2013a} finds a median $c=7$ in a sample of local galaxies.
Thus, we shall assume that the {\it local} columns ($\Sigma_{\rm CO/H_2}$) in the COLDGASS sample are higher by such a factor (adopting $c=5$).
Another implicit assumption in Eq.~\ref{eq:f_C+_H2} is that when observing perpendicular to the galaxy disc we typically have up to one cloud in the line of sight, as opposed to several clouds behind each other.
This is a reasonable assumption when most of the molecular gas is in large clouds that occupy a sizable fraction of the disc scale-height.

We should also consider clouds with insufficient columns to reach the max C$^+$ column and form a CO core - can such clouds hold a significant fraction of the molecular gas in a galaxy?
Let us assume an extreme case in which all the area of the galaxy that has no CO column in it is filled with the maximal $N_{\rm C^+/H_2}$ column.
In this special case we do not need to account for the clumpiness factor since $\Sigma_{\rm C^+/H_2}$ is now uniform across the disc and the observed $\Sigma_{\rm CO/H_2}$ is already averaged over the disc.
Therefore, setting $c=1$ in Eq.~\ref{eq:f_C+_H2} gives the upper limit to $f_{\rm C^+/H_2}$.

For the median CO surface densities in the local massive galaxies of the COLDGASS sample we get $f_{\rm C^+/H_2} \approx 0.4$ and an upper limit ($c=1$) of $f_{\rm C^+/H_2} < 0.77$
In these galaxies the ${\rm C^+/H_2}$ gas can reach a significant fraction of the total molecular gas due to the low surface densities.
The degree of clumpiness in these galaxies can have a major effect on $f_{\rm C^+/H_2}$.
Also, in case of a large degree of clumpiness the local galaxies can {\it potentially} hold a large amount of ${\rm C^+/H_2}$ gas in small clouds with no CO, between the GMCs.

W10 calculated the ${\rm C^+/H_2}$ fraction assuming clouds of $10^6$~M$_\odot$ in mass and a spherical geometry.
Their numerical model which is intended to represent the typical GMCs in local galaxies yields a typical $f_{\rm C^+/H_2} \approx 0.3$, close to our estimation above.
In the Milky Way, \citet{Pineda2013} and \citet{Langer2014} derived a global $f_{\rm C^+/H_2} \approx 0.3$ and $f_{\rm C^+/H_2} \approx 0.4$ (respectively) from  [C~{\sevensize II}] mapping with {\it Herschel}.
Thus, both the W10 theoretical model and Milky Way measurements are similar to the typical $f_{\rm C^+/H_2}$ that we derive for the local massive spiral galaxies from the COLDGASS survey.

At higher redshifts ($1<z<2.5$), the PHIBBS survey observed typical star forming galaxies with a median mass of $6\times10^{10}$~M$_\odot$.
The median molecular gas surface density is $\Sigma_{\rm H_2}=560$~M$_\odot$~pc$^{-2}$ and the survey reaches molecular surface mass densities down to $\Sigma_{\rm H_2}\sim$100~M$_\odot$~pc$^{-2}$.
The expected metallicity for the PHIBBS from the mass--metallicity relation \citep{Erb2006a} is $\sim0.7$, thus $N_{\rm C^+/H_2}$ increases by a factor of $Z^{\prime-1}=1.4$ over the expected value at $Z^\prime=1$, reaching $\Sigma_{\rm C^+/H_2} \approx 95$~M$_\odot$~pc$^{-2}$.
Adopting an ad-hoc $c=5$ clumpiness factor similar to the low redshift galaxies, the CO-dark gas fractions are $f_{\rm C^+/H_2}=0.03$ for the median PHIBBS galaxy, and the upper limit ($c=1$) is $f_{\rm C^+/H_2} < 0.11$. 
Even for the lowest CO surface brightness galaxies in the PHIBBS sample $f_{\rm C^+/H_2}=0.16$ is still quite low.

The gas surface densities of the high redshift galaxies are so high that even if the metallicity is as low as $Z^\prime=0.1$, $f_{\rm C^+/H_2}$ will not be very high.
Going to even lower $Z^\prime \approx 0.01$, the ${\rm C^+/H_2}$ columns saturate at $N_{\rm C^+/H_2}=2\times10^{23}$~cm$^{-2}$ (Figure~\ref{fg:N_CpH2_vs_Z}), which translates to $\Sigma_{\rm C^+/H_2}=2000$~M$_\odot$~pc$^{-2}$ two-sided column.
Such values are comparable to the local molecular surface densities measured from CO in PHIBBS galaxies (after applying the clumpiness factor), though the comparison is not strictly valid since the PHIBBS galaxies have metallicities of $Z^\prime>>0.01$.

One must keep in mind that CO at low-$Z^\prime$ is sensitive to the presence of CR.
With the high global SFRs at redshifts $z>1$ one may expect a high flux of CR that can dramatically lower the CO abundance at the dense cores of low metallicity clouds.
An example of this can be seen in the right column of Figure~\ref{fg:abundance_structure_combined} that assumes quite a moderate CR ionization rate.
Our setup for the numerical calculation is not intended to simulate the CO cores, and thus temperatures are likely too high while densities too low, but the general behavior is still valid.
In such extreme cases, even the core will become `CO-dark' and dominated by neutral C (and also some C$^+$) due to cosmic rays ionizing H$_2$ and dissociating the CO via induced processes.
This effect should not be confused with the processes in the ${\rm C^+/H_2}$ envelope which is devoid of CO due to dissociation by the external UV field.

%%%%%%%%%%%%%%%%%%%%%%%%%%%%%%%%%%%%
%
\section{Conclusions} \label{sec:conclusions}
%
%%%%%%%%%%%%%%%%%%%%%%%%%%%%%%%%%%%%
In this paper we have studied the transition of carbon from C$^+$ to C and CO in the photon dominated regions of molecular clouds.
We derived an analytic expression for $N_{\rm C^+}^{\rm tot}$ the total column of C$^+$, which is a function of the metallicity, the ratio $I_{\rm UV}/n$, and a weak implicit function of the temperature.
We numerically calculated the abundance structures using the MEUDON PDR code and find an excellent agreement with our analytic $N_{\rm C^+}^{\rm tot}$.

The properties of the C$^+$ layer in the PDRs of molecular clouds can be summarized as follows:
\begin{enumerate}

 \item The transition from C$^+$ to C or CO as the dominant form of carbon occurs at $A_{\rm V}=1.5$ for metallicities of $Z^\prime>0.1$. The total optical depth associated with the total C$^+$ column is $A_{\rm V}=2$, consistent with many previous results.
 However, at $Z^\prime<0.1$ this transition happens at lower $A_{\rm V}$ due to H$_2$ shielding that becomes more significant than dust shielding.
 
 \item At all relevant conditions the column of C$^+$ inside the molecular H$_2$ region is significantly larger than the C$^+$ column in the atomic \HI\ region.
 We can thus treat the C$^+$ as if originating entirely from the H$_2$ region and $N_{\rm C^+/H_2} \approx N_{\rm C^+}^{\rm tot}/(AZ^\prime)$ is the column of the `CO-dark' gas.
 
 \item $N_{\rm C^+/H_2}$ the total molecular gas column associated with C$^+$  is mostly a function of $Z^\prime$ and only a logarithmic function of $I_{\rm UV}/n$.
 If adopting the CNM relation, $I_{\rm UV}/n$ ratio itself becomes a function of the metallicity and thus $N_{\rm C^+/H_2}$ depends on $Z^\prime$ alone.
 
 \item $N_{\rm C^+/H_2}$ is linear with $Z^{\prime-1}$ for $Z^\prime \gtrsim 0.1$.
 At solar metallicity, the typical column of the \CpHt\ `CO-dark' molecular gas on the face of the cloud is $N_{\rm C^+/H_2}=3\times10^{21}$~cm$^{-2}$.
 At $Z^\prime \lesssim 0.1$ H$_2$ shielding limits the gas column associated with C$^+$, that saturates when $Z^\prime \approx 0.01$, at a column of $N_{\rm C^+/H_2}=10^{23}$~cm$^{-2}$.
 
 \item the $N_{\rm C^+/H_2}$ columns we derive are in practice upper limits that are reached only if the total gas columns are large enough ($N > 2\times N_{\rm C^+/H_2}$).
 Smaller clouds will not develop a CO core in spite of containing molecular gas.
 Total columns in excess of $2 \times N_{\rm C^+/H_2}$ will add directly to the CO column, thus the ${\rm C^+/H_2}$ fraction out of the molecular gas depends on the total column of the cloud and $Z^\prime$ (therefore $A_{\rm V}$ of the cloud).
 This is in agreement with the numerical model of W10 in spite of their model assuming a spherical geometry and a constant CO mass.
 
 \item At metallicities $0.1<Z^\prime<1$, C$^+$ changes into C and CO without forming a significant neutral C layer.
 At $Z<0.1$ a significant neutral C column can exist.
 However, most of the neutral C is mixed with the CO and only a small fraction is in the C$^+$ region or form a neutral C dominated region and contribute to the `CO-dark' molecular gas.
 
 \end{enumerate}

The application of our analytic $N_{\rm C^+/H_2}(Z^\prime)$ to entire galaxies is not straight forward.
However, under the assumption that most of the molecular gas is in clouds large enough to include a CO core we estimate the typical ${\rm C^+/H_2}$ fraction in local massive spirals to be $f_{\rm C^+/H_2} \approx 0.4$.
This result is consistent with spectroscopic measurements in the Milky Way.
In $1<z<2.5$ normal star forming galaxies the molecular gas columns observed in CO are so high that the ${\rm C^+/H_2}$ gas cannot hold a significant fraction of the molecular gas (typical $f_{\rm C^+/H_2} \lesssim 0.11$).

 [C~{\sevensize II}]~158~$\mu$m observations are a promising tool to directly observe the ${\rm C^+/H_2}$ gas.
However, the conversion from  [C~{\sevensize II}]~158~$\mu$m brightness to gas mass in PDRs depends on estimations of the temperatures, densities and the fraction of  [C~{\sevensize II}] emitted from other phases of the ISM.
It is important to emphasize that our prediction of $N_{\rm C^+/H_2}$ depends (almost) only on $Z^\prime$ and is independent of the emitted  [C~{\sevensize II}].

%%%%%%%%%%%%%%%%%%%%%%%%%%%%%%%%%%%%%%%%%%%%%%%%%%
%%%%%%%%%%%%%%%%%%%%%%%%%%%%%%%%%%%%%%%%%%%%%%%%%%
\section*{Acknowledgments}
We thank Reinhard Genzel, Frank Le Petit, Evelyne Roueff, and
Linda Tacconi for helpful discussions.
We thank the referee for the helpful comments that improved our paper.
This work was supported by the DFG via German-Israeli Project Cooperation grant STE1869/1-1/GE625/15-1, and by a PBC
Israel Science Foundation I-CORE Program, grant 1829/12.

%%%%%%%%%%%%%%%%%%%%%%%%%%%%%%%%%%%%%%%%%%%%
\bibliographystyle{mnras}
\bibliography{bibli}
%%%%%%%%%%%%%%%%%%%%%%%%%%%%%%%%%%%%%%%%%%%%%
% latex x bibtex x latex x latex x

%%%%%%%%%%%%%%%%%%%%%%%%%
%  FIGURES   %%
%%%%%%%%%%%%%%%%%%%%%%%%%

%%%%%%%%%%%%%%%%%%%%%%%%%%%%%%%%%%%%%%%%%%%%%%
%
\appendix
%
%%%%%%%%%%%%%%%%%%%%%%%%%%%%%%%%%%%%%%%%%%%%%%

%%%%%%%%%%%%%%%%%%%%%%%%%%%%%%%%%%%%%%%%%%%%%%%%%%%%%%%%%%%%
\section{$N_{\rm C^+}$ with a column-dependent $\sigma_{\rm H_2}$}
\label{app:NCp_1st_order_correction}
%%%%%%%%%%%%%%%%%%%%%%%%%%%%%%%%%%%%%%%%%%%%%%%%%%%%%%%%%%%%%
We repeat the calculation of $N_{\rm C^+}^{\rm tot}$ from Section~\ref{sec:analytic_C+_column}, but in this case  $\sigma_{\rm H2}$ is not a constant.
The shielding of carbon by H$_2$ lines does not operate in the same way as the dust.
While dust is not very sensitive to the exact wavelength of the UV photons, H$_2$ shielding is done through spectral absorption lines.
Therefore the cross-section for interaction for the photons close in wavelength to the line centers is much higher than for photons at wavelengths between the lines.
The former are quickly depleted, and as we go deeper into the cloud photons farther away from the line centers start to get absorbed in the line wings.
Thus, the {\it effective} cross-section for absorption gets lower with increasing H$_2$ column.

The effect of diminishing H$_2$ absorption cross section is moderated by the density of the H$_2$ lines.
Wavelengths farther away from the center of one line means closer to the next line.
When the line wings overlap and the surviving UV photons are far from the line centers, the combined absorption cross section of all the H$_2$ lines becomes a weaker function of the total column.
For this reason, while not perfectly accurate, H$_2$ shielding can be approximated as a constant cross section absorber, as we did in \S~\ref{sec:H2_line_blocking} and as done in previous works \citep[e.g.,][]{deJong1980, Tielens1985a}

The dust competes with H$_2$ lines in absorbing the UV photons and absorbs photons close to line centers, as well as between the lines indiscriminately.
Thus, the effect of dust is to increase the mean cross section for H$_2$ absorption at a given depth in the cloud by decreasing the population of photons between the H$_2$ lines (i.e., increase the relative weight of the near-line photons in the effective $\sigma_{\rm H2}$).

\begin{figure}
 \centering
 \includegraphics[width=0.5\textwidth,keepaspectratio=true]{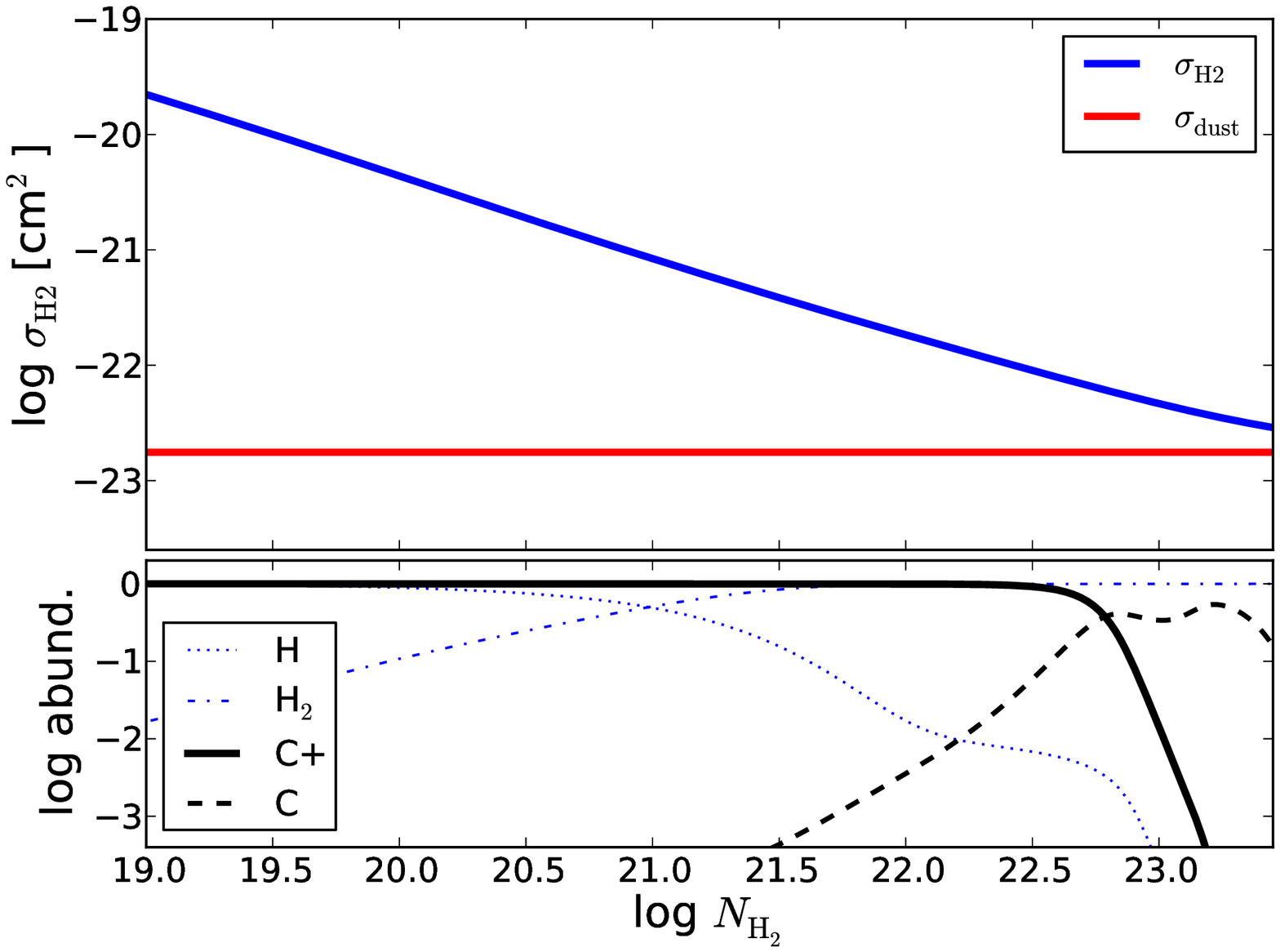}
 % sigma_H2_vs_N_H2.eps: 0x0 pixel, 300dpi, 0.00x0.00 cm, bb=13 175 598 616
 \caption{The effective $\sigma_{\rm H2}$ as a function of the H$_2$ column for a model with $Z^\prime=0.01$, $I_{\rm UV}=10$ and $n=590$ cm$^{-3}$. The bottom panel shows the specie relative abundances as a function of the H$_2$ column, using the same x-axis scale as the top panel.}
 \label{fg:sigma_H2_vs_N_H2}
\end{figure}

We can use the output of the MEUDON PDR code that tracks the photoionization rate of C to calculate the effective cross section for absorption by H$_2$ as a function of the total H$_2$ column $N_{\rm H_2}$.
The effective $\sigma_{\rm H2}$ at a given H$_2$ column in the cloud is defined by the C$^+$ ionization rate, such that:
\begin{equation}
 R(N_{H_2}) = R(0) \exp(-\sigma_{\rm g,\odot}Z^\prime N - 2\sigma_{\rm H_2}N_{\rm H_2})
\end{equation}
where $R(0)$ is the carbon ionization rate at the face of the cloud.
The photoionization rate as a function of depth is one of the outputs of the code.
Such an example is plotted in Figure~\ref{fg:sigma_H2_vs_N_H2} and we can see that $\sigma_{\rm H2}$ can be well described by a power-law.
\begin{equation}
 \sigma_{\rm H_2} = b^\prime N_{\rm H_2}^\alpha \approx b(N_{\rm C+}/(AZ^\prime))^\alpha
 \label{eq:App_sigmaH2_powerlaw}
\end{equation}
We are interested in $\sigma_{\rm H_2}$ close to the maximal depth of the C$^+$ dominated region, where the C$^+$ abundance falls.
At that depth, most of the total hydrogen column is in H$_2$ and almost all of the carbon column is in C$^+$, thus we can use  $N_{\rm C^+}/(AZ^\prime)$ instead of $N_{\rm H_2}$ in the power-law, with a minor adjustment to the coefficients.
The coefficients change only little depending on the conditions.

We now repeat the calculation from Section~\ref{sec:analytic_C+_column} till we get to the integration (similar to Eq.~\ref{eq:C+_integration}) with the added $\sigma_{\rm H2}$:
\begin{equation}
 \displaystyle\int\displaylimits_0^{{N}_{\rm C^+}^{\rm tot}} \exp\left\{ +\frac{\sigma_{\rm g,\odot}+\sigma_{\rm H_2}/Z^\prime}{A}{N}_{\rm C^+} \right\} d{N}_{\rm C^+} =
\mathcal{Y} \displaystyle\int\displaylimits_0^{\infty} \exp \left\{ -\frac{\sigma_{\rm g,\odot}+\sigma_{\rm H_2}/Z^\prime}{A}{N}_{\rm C} \right\} d{N}_{\rm C}
 \label{eq:App_C+_integration}
\end{equation}
$\sigma_{\rm H_2}$ is not constant, but since the left-hand side of Eq.~\ref{eq:App_C+_integration} is a rising exponent, the significant values of $\sigma_{\rm H2}$ are evaluated near the upper limit of the integration, i.e., most of the integrated total comes from integrating the kernel close to the upper limit.

Let us linearize the term $\sigma_{\rm H_2}N_{C^+}$ near the upper limit of the integration by
doing a Tylor expansion around $N_{C^+}=N_0$, omitting terms of the order of 2 and above, and using Eq.~\ref{eq:App_sigmaH2_powerlaw} to express $\sigma_{\rm H_2}$:
\begin{eqnarray}
 \sigma_{\rm H_2}N_{C^+} = \frac{b}{(AZ^\prime)^{\alpha}}N_{C^+}^{\alpha+1} \approx
 \frac{b}{(AZ^\prime)^{\alpha}}N_{0}^{\alpha+1} + \frac{b(\alpha+1)}{(AZ^\prime)^{\alpha}}N_0^{\alpha}(N_{C^+}-N_0) = \nonumber \\
 = -\alpha b \left(\frac{N_0}{AZ^\prime}\right)^\alpha N_0 + (\alpha+1)b \left(\frac{N_0}{AZ^\prime}\right)^\alpha N_{C^+}
 \label{eq:taylor_expand}
\end{eqnarray}

We can now use the expression in Eq.~\ref{eq:taylor_expand} inside the right hand side of the integral in Eq.~\ref{eq:App_C+_integration}.
$N_0$ is a constant representing a C$^+$ column close to $N_{C^+}^{tot}$.
The dynamic range of $N_{C^+}^{tot}$ is quite limited and is always of the order of $1\times10^{17}$~cm$^{-2}$, even for large variations in the radiation, density and metallicity.
Therefore, we will fix $N_0$ to this value.
On the right hand side of Eq.~\ref{eq:App_C+_integration} , we will take $\sigma_{\rm H_2} = b(N_0/(AZ^\prime))^\alpha$. The reason is that this kernal is a decaying exponent and C column rises sharply where C+ abundance drops sharply, meaning around a C$^+$ column of the same order as $N_0$.
The solution is:
\begin{equation}
 N_{C^+}^{tot} = \frac{A}{\sigma_{g,\odot} + (\alpha+1)b(N_0/A)^\alpha Z^{\prime-\alpha-1} } \ln\left( \mathcal{Y}\mathcal{T} + 1 \right)
 \label{eq:App_NCp_tot}
\end{equation}
where $\mathcal{T}$ is defined as:
\begin{equation}
 \mathcal{T} \equiv \left( 1 + \frac{\alpha}{\sigma_{\rm g,\odot} b^{-1} (N_0/A)^{-\alpha} Z^{\prime\alpha+1} + 1}\right) \exp\left(\alpha b \left(\frac{N_0}{AZ^\prime}\right)^{\alpha+1}\right)
\end{equation}
When comparing with the previous result for a constant $\sigma_{\rm H_2}$ (Eq.~\ref{eq:N_C+_tot}), we now have an additional term inside the log, $\mathcal{T}$, that is a function of metallicity only and the ad hoc $N_0=1\times10^{17}$~cm$^{-2}$ parameter we set.
We find that setting $\alpha=-0.5$ and $\log(b)=-10.73$ yields a good compromise that can fit the entire library of numerical models we produced within $\pm0.1$~dex accuracy.
For the range of $10^{-2}<Z^\prime<3$, in the definition of $\mathcal{T}$ the expression in parenthesis outside the exponent evaluates to be in the range of 0.85--1.0 and hence can be omitted.
If we use the explicit values for $AZ^\prime$ and $\sigma_{\rm g,\odot}$, the final expression for $N_{C^+}^{tot}$ is (equivalent to Eq.~\ref{eq:N_C+_tot}):
\begin{equation}
 N_{\rm C^+}^{\rm tot} = \frac{7.5\times10^{16}\,{\rm cm^{-2}}}{1 + 0.19Z^{\prime-0.5}} \ln\left( \mathcal{Y}e^{-0.26Z^{\prime-0.5}} + 1 \right)
 \label{eq:App_NCp_tot_final}
\end{equation}
and the total gas column is (equivalent to Eq.~\ref{eq:N_C+_total_H}):
\begin{equation}
 N_{\rm (HI+H_2),C^+}^{\rm tot} = \frac{5.7\times10^{20}\,{\rm cm^{-2}}}{Z^\prime + 0.19Z^{\prime 0.5}} \ln\left( \mathcal{Y}e^{-0.26Z^{\prime-0.5}} + 1 \right)
 \label{eq:App_NCp_tot_H_final}
\end{equation}

\begin{figure}
 \centering
 \includegraphics[width=0.5\textwidth]{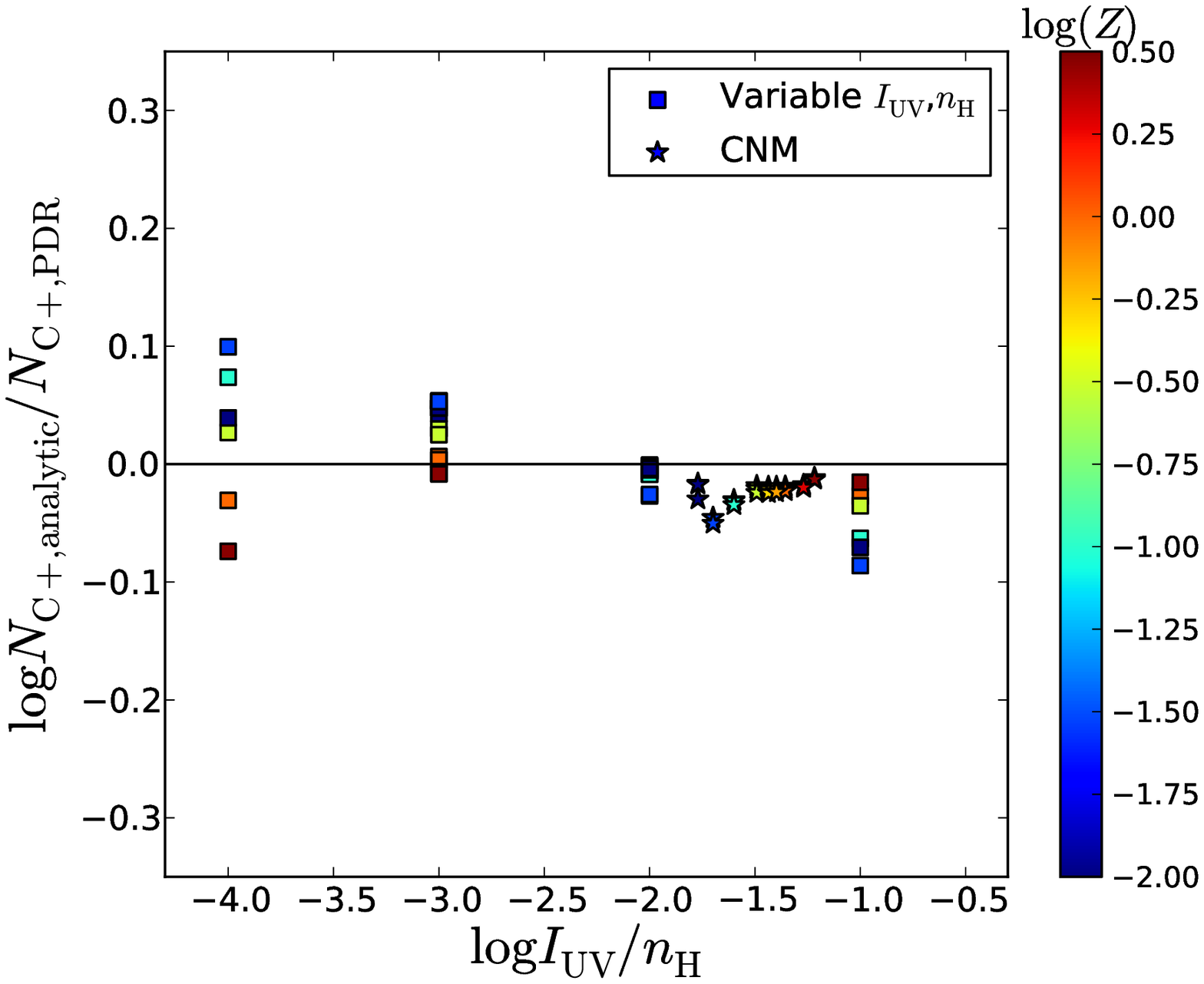}
 % N_Cp_accuracy_1st_order_tauH2.eps: 0x0 pixel, 300dpi, 0.00x0.00 cm, bb=13 175 598 616
 \caption{The ratio of the C$^+$ column as calculated by the analytic expression to the result of the numerical calculation, as a function of $I_{\rm UV}/n$. This figure is similar to Figure~\ref{fg:NCp_analyt_Meudon} except for using Eq.~\ref{eq:App_NCp_tot_final} instead of Eq.~\ref{eq:N_C+_tot} in $N_{\rm C^+, analytic}$.}
 \label{fg:App_NCp_1storder_accuracy}
\end{figure}

In Figure~\ref{fg:App_NCp_1storder_accuracy} we compare $N_{C^+}^{tot}$ as predicted by Eq.~\ref{eq:App_NCp_tot_final} with the results of the Meudon code for our models grid.
The analytic expression is accurate within 0.1~dex with no significant trend with $I_{\rm UV}/n$.
The improvement over Eq.~\ref{eq:N_C+_tot} that assumes a constant $\sigma_{\rm H2}$ and plotted in Figure~\ref{fg:NCp_analyt_Meudon} is clear.

In this paper we adopt a constant $\sigma_{\rm H_2}$ in spite of its lower accuracy.
The simpler, cleaner expressions in Eq.~\ref{eq:N_C+_tot} and Eq.~\ref{eq:N_C+_total_H} offer a better intuitive understanding of their behaviors.
We feel that for the most plausible cases (near CNM conditions) a constant $\sigma_{\rm H_2}$ is accurate enough and even at more extreme $I_{\rm UV}/n$, the lower accuracy is still of the same scale or better than other systematic uncertainties encountered in practical applications of the theory.

%%%%%%%%%%%%%%%%%%%%%%%%%%%%%%%%%%%%%%%%%%%%%%%%%%%%%%%%%%%%%%%%%%%%%%%%%%%%%%%%%
%%%%%%%%%%%%%%%%%%%%%%%%%%%%%%%%%%%%%%%%%%%%%%%%%%%%%%%%%%%%%%%%%%%%%%%%%%%%%%%%%
%%%%%%%%%%%%%%%%%%%%%%%%%%%%%%%%%%%%%%%%%%%%%%%%%%%%%%%%%%%%%%%%%%%%%%%%%%%%%%%%%

\end{document}